%% file: main.tex
\documentclass[10pt,onecolumn,a4paper]{article}

\input{preamble}

\usepackage[margin=1.25in]{geometry}
\usepackage{footmisc}

\begin{document}
	

	\title{Design, Implementation, and Automation\\ of a Risk Management Approach\\ for Man-at-the-End Software Protection\footnote{Preprint submitted to Computers \& Security.}}
	
	
	\author{Cataldo Basile\footnote{Cataldo Basile and Bjorn De Sutter share dual first authorship.}\\Politecnico di Torino\\cataldo.basile@polito.it\\
		\and
		Bjorn De Sutter\textsuperscript{$\dagger$}\\Ghent University\\bjorn.desutter@ugent.be\\
		\and
		Daniele Canavese\\Politecnico di Torino\\daniele.cavanese@polito.it\\
		\and
		Leonardo Regano\\Politecnico di Torino\\leonardo.regano@polito.it\\
		\and
		Bart Coppens\\Ghent University\\bart.coppens@ugent.be
	}
	
	
	
	
	

	\maketitle
	
	\begin{abstract}
		The last years have seen an increase in Man-at-the-End (MATE) attacks against software applications, both in number and severity.
		However, software protection, which aims at mitigating MATE attacks, is dominated by fuzzy concepts and security-through-obscurity.
		This paper presents a rationale for adopting and standardizing the protection of software as a risk management process according to the NIST SP800-39 approach.
		We examine the relevant constructs, models, and methods needed for formalizing and automating the activities in this process in the context of MATE software protection. We highlight the open issues that the research community still has to address. We discuss the benefits that such an approach can bring to all stakeholders.
		In addition, we present a Proof of Concept (PoC) decision support system that instantiates many of the discussed construct, models, and methods and automates many activities in the risk analysis methodology for the protection of software.
		Despite being a prototype, the PoC's validation with industry experts indicated that several aspects of the proposed risk management process can already be formalized and automated with our existing toolbox and that it can actually assist decision making in industrially relevant settings.
	\end{abstract}
	
	


	\section{Introduction}
	\label{sec:intro}
	\input{intro_aldo}

	\input{approach}

	\input{rationale}

	
	\input{requirements}

	\input{workflow}

	
	


	\input{conclusion}
	
	\section{Funding}
	This research was partly funded by the Cybersecurity Initiative Flanders (CIF) from the Flemish Government and by the Fund for Scientific Research - Flanders (FWO) [Project No.~3G0E2318]. Part of the presented results were obtained in the context of the ASPIRE FP7 research project. This project ran until October 2016 and has received funding from the European Union Seventh Framework Programme (FP7/2007-2013) under grant agreement number 609734.
	

	\appendix
	
	\section{The questionnaire for the expert assessment of the \esp. }
	\label{sec:appendix}
	
	\changed{This questionnaire is semantically equivalent to the questionnaire provided to the experts during the ASPIRE project. We  removed ASPIRE-specific terms, which have been substituted with a wording  coherent with this paper, and slightly rephrased some sentences for clarity.}
	%
		\begin{tcolorbox}[colback=white,breakable, sharp corners]
			\small
			\textbf{Preparation}
			
			The primary Assets are the ones selected and protected for preparing the application for the tiger team experiments. The assets' relevance has been assigned by \textit{NAME} during the \textit{PLACE} meeting.
			
			\textit{
				NOTE: In general, any feedback on the visualisation, presentation, data, and options in both the tool and the report is welcome.}

			\textbf{Application Parts}
			\begin{enumerate}
				\item Is the list of the APPLICATION PARTS \changed{useful} for your software protection purposes/tasks?
				\item Report if you want to see them differently or if some information is missing.
			\end{enumerate}

			\textbf{Attack Steps}
			\begin{enumerate}
				\item VALIDATE if the attack steps identified by the automatic analysis are meaningful. Check if they are exhaustive.
				
				\item For all the attack steps, VALIDATE if the \textit{Suggested Protections} are correct, sound, proper, and effective. Report if the estimated attack EFFECTIVENESS is correct.
				
				\item REPORT if the attack data shown are useful. Report if you want to see them differently or if you miss some important information.
			\end{enumerate}
			
			\textbf{Golden Configurations}
			\begin{enumerate}
				\item VALIDATE if the golden combinations are meaningful and you consider them effective/optimal. 
				\item For all the 10 golden combinations presented in the tool and report, you should NOTIFY us if you noticed something strange in the use of the protection techniques, like some association/combination of techniques which is strange according to your expert judgment, the use of some techniques to protect specific assets that you may consider anomalous
				\item REPORT us if the data shown are useful. Report if you want to see them differently or if you miss some important information.
			\end{enumerate}
			
			\textbf{ Asset hiding }
			
			\begin{enumerate}
				\item VALIDATE if the layer-two protections (asset hiding) and check if they are effective/optimal. 
				\item NOTIFY if you noticed something anomalous in the use of techniques, e.g., some association of techniques which is strange for experts, the use of techniques for assets in anomalous/inappropriate ways, or some cases where you wouldn't extend/randomly pick other areas.
				\item REPORT us if the data shown are useful. Report if you want to see them differently or if you miss some important information.
				\item REPORT if the estimated maximum degradation thresholds (overheads sections of the Golden Configurations and Asset Hiding) are reasonable. 
			\end{enumerate}
		\end{tcolorbox}
		\parpic{\includegraphics[width=1in,clip,keepaspectratio]{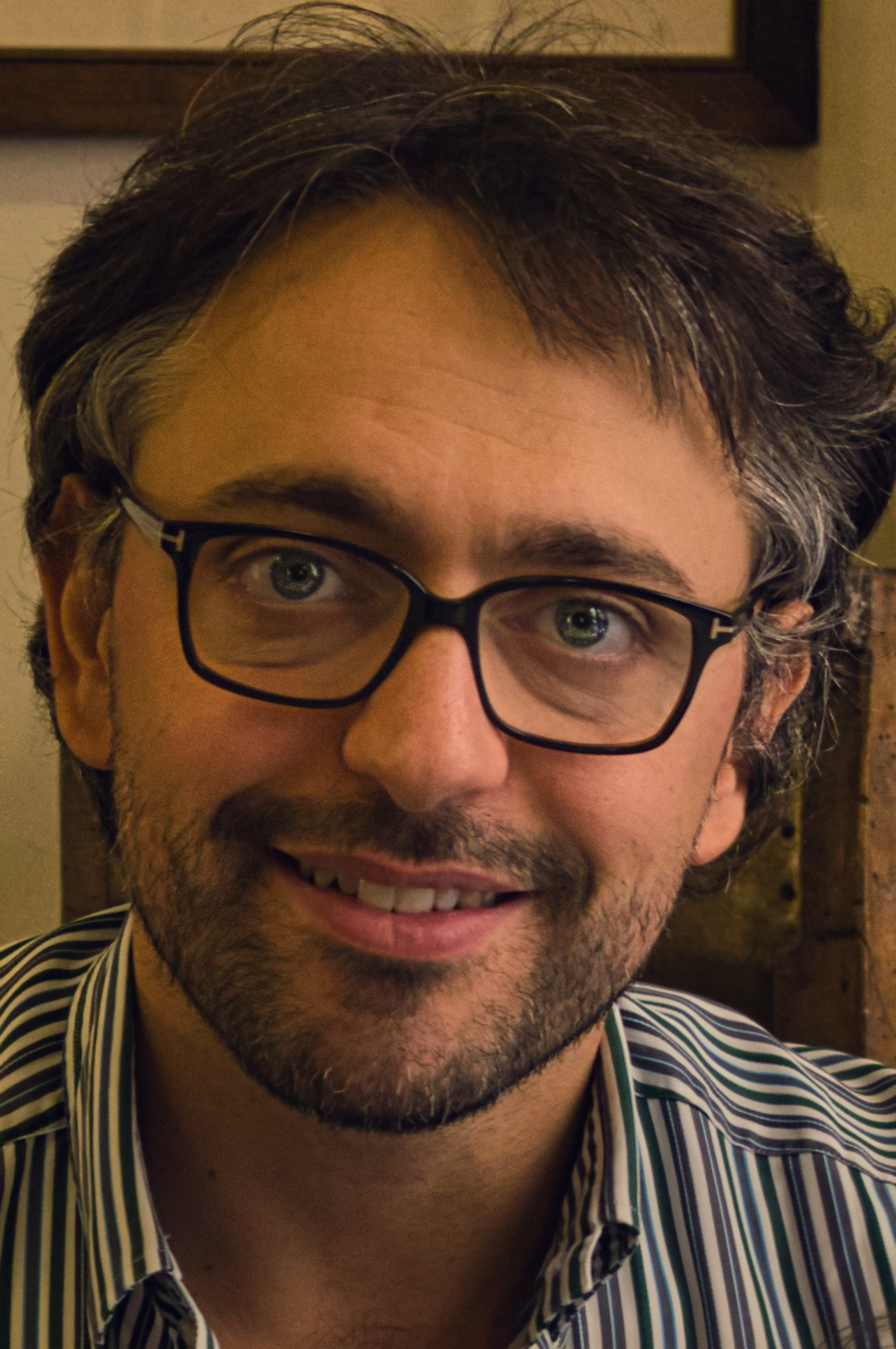}}
		\noindent\textbf{Cataldo Basile} is an assistant professor at the Politecnico di Torino, from which he received an M.Sc.in 2001 and a Ph.D. in Computer Engineering in 2005. His research concerns software protection, software attestation, policy-based security management, and general models for detecting, resolving, and reconciling security policy conflicts.
		\vspace{5em}
		
		\parpic{\includegraphics[width=1in,clip,keepaspectratio]{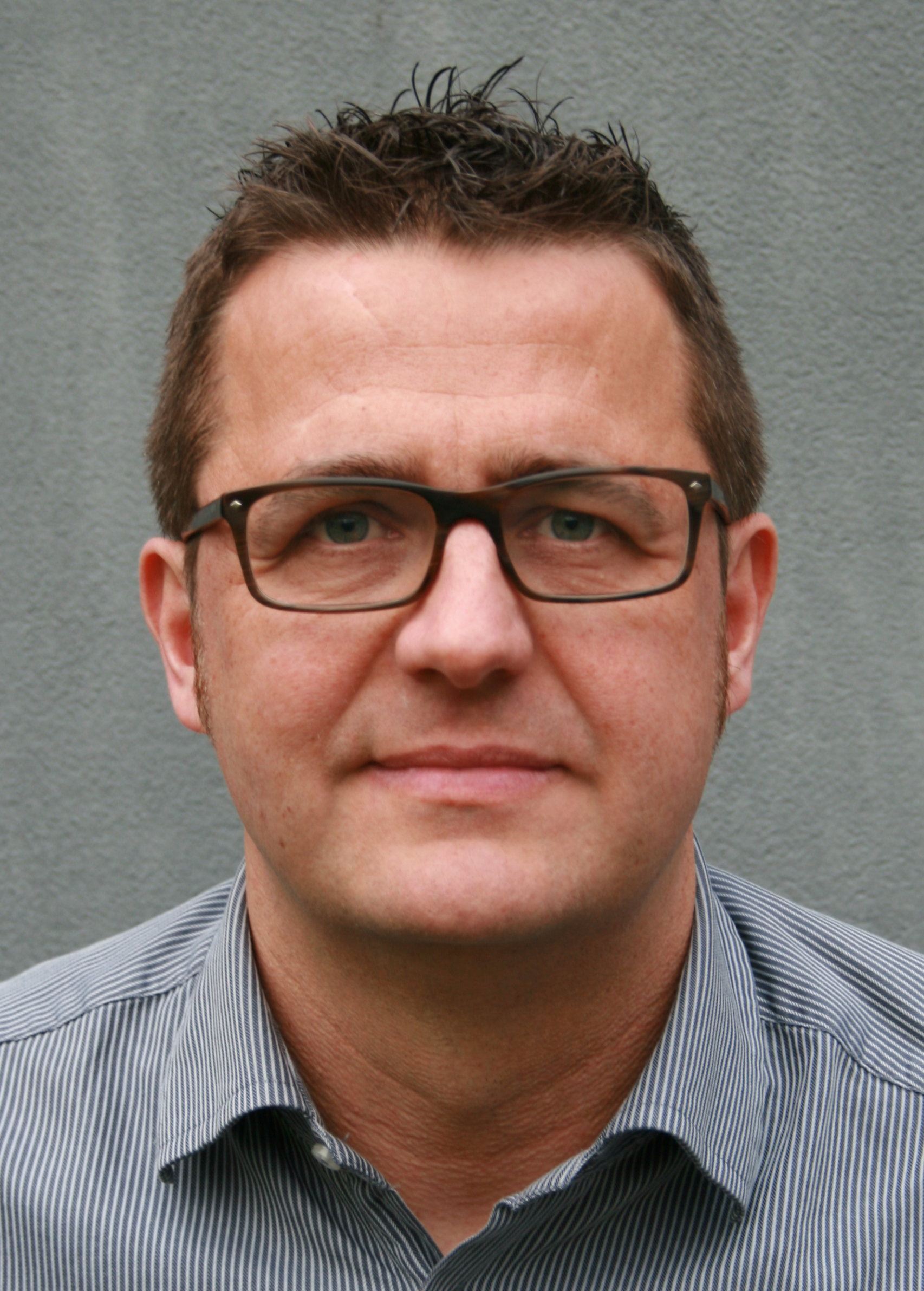}}
		\noindent\textbf{Bjorn De Sutter} is associate professor at Ghent University in the Computer Systems Lab. He obtained his MSc and PhD degrees in Computer Science from the university's Faculty of Engineering in 1997 and 2002. His research focuses on techniques to aid programmers with non-functional aspects such as performance and software protection to mitigate reverse engineering, software tampering, code reuse attacks, fault injection, and side channel attacks. He co-authored over 100 papers.
		\vspace{2em}
		
		\parpic{\includegraphics[width=1in,clip,keepaspectratio]{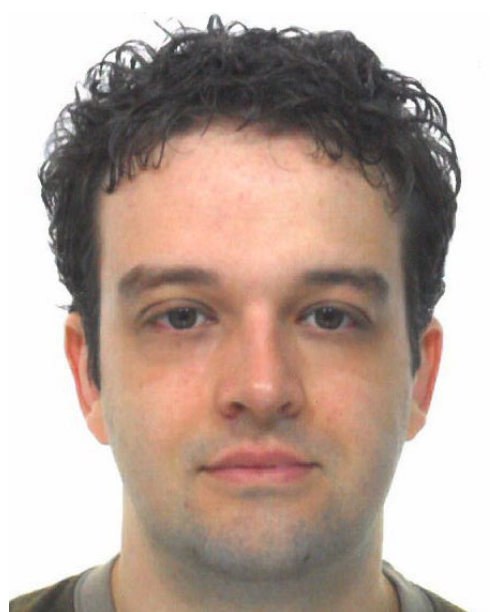}}
		\noindent\textbf{Daniele Canavese} received an M.Sc. degree in 2010 and a Ph.D. in Computer Engineering in 2016 from Politecnico di Torino, where he is currently a research assistant. His research interests are concerned with security management via machine learning and inferential frameworks, software protection systems, public-key cryptography, and models for network analysis.
		\vspace{3em}
		
		\parpic{\includegraphics[width=1in,clip,keepaspectratio]{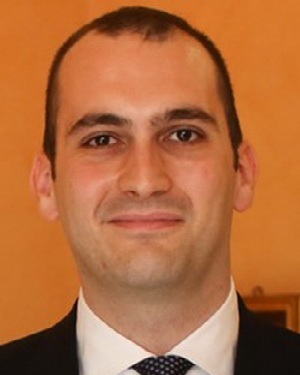}}
		\noindent\textbf{Leonardo Regano}  received an M.Sc.degree in 2015 and a Ph.D. in Computer Engineering in 2019 from Politecnico di Torino, where he is currently a research assistant. His current research interests focus on software security, artificial intelligence and machine learning applications to cybersecurity, security policies analysis, and software protection techniques assessment.
		\vspace{3em}

		\parpic{\includegraphics[width=1in,clip,keepaspectratio]{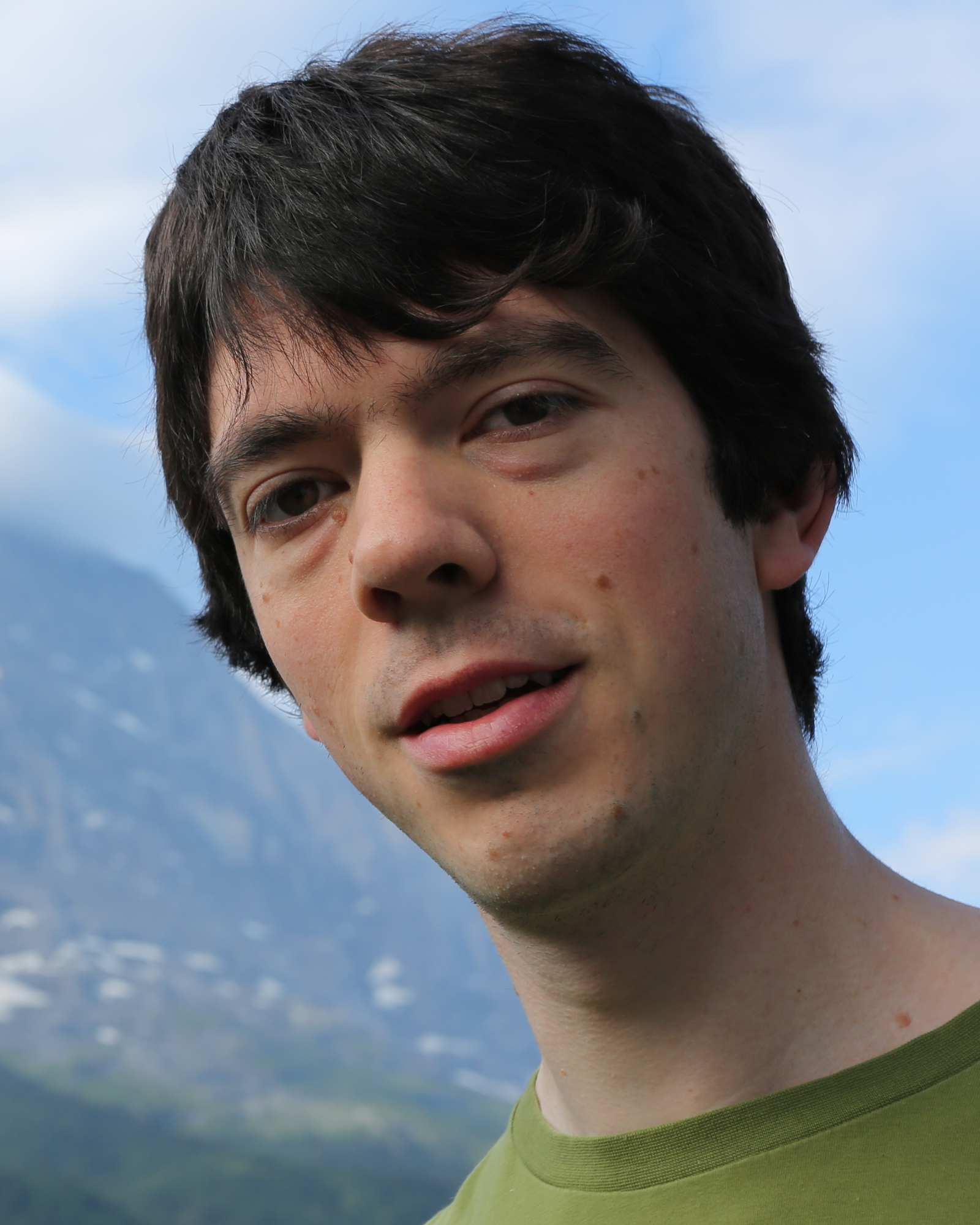}}
		\noindent\textbf{Bart Coppens} is a assistant professor at Ghent University in the Computer Systems Lab. He received his PhD in Computer Science Engineering from the Faculty of Engineering and Architecture at Ghent University in 2013. His research focuses on protecting software against different forms of attacks using compiler-based techniques and run-time techniques.

	\end{document}

%% file: intro_aldo.tex


In the \mate attack model, attackers have white-box access to the software. This means they have full control over the systems on which they identify successful attack vectors in their lab, for which they use all kinds of attacker tools such as simulators, debuggers, disassemblers, decompilers, etc. Their goal is to reverse engineer the software (e.g., to steal valuable secret algorithms or embedded cryptographic keys or to find vulnerabilities in the code), to tamper with the software (e.g., to bypass license checks or to cheat in games), or to execute it in unauthorized ways (e.g., run multiple copies in parallel). In general, \mate attacks target software to violate the security requirements of assets present in that software.

{\mate} {\softprot} then refers to \emph{protections deployed within that software} to mitigate \mate attacks.\footnote{For the sake of brevity, we will omit the {\mate} and simply use {\softprot} to mean {\mate} {\softprot} in the remainder of this paper.}  {\softprot} is hence much narrower than the broad umbrella of software security. The latter also includes scenarios in which software is exploited to violate \emph{security requirements of other system components}, e.g., infiltrating networks or escalating privileges. In such scenarios, attackers start with limited capabilities, such as having only unprivileged, remote access to a computer via a web server interface. 

Because \mate attackers have full control over the devices in their labs, \softprot needs to defend assets in the software \emph{without relying on external services} running on those devices. Instead, defenders can only rely upon protections deployed within the protected software or remote servers. 
Advances in cryptography have yielded techniques that provide strong security guarantees but also introduce orders of magnitude performance overhead~\cite{horvath2020cryptographic}.
They are hence rarely practical today. Then again, practical {\softprot} is still dominated by fuzzy concepts and techniques~\cite{collbergbook}. {\softprot}s such as remote attestation, obfuscation, and anti-debugging do not aim to mitigate \mate attacks completely.
Instead, they aim to delay attacks and put off potential attackers by increasing the expected cost of attacks and by decreasing the expected \roi. 

As observed during a recent Dagstuhl seminar on \softprot Decision Support and Evaluation Methodologies~\cite{Dagstuhl}, the \softprot field is facing severe challenges: \sto is omnipresent in the industry; 
{\softprot} tools and consultancy are expensive and opaque; 
there is no generally accepted method for evaluating {\softprot}s and {\softprot} tools. Moreover, {\softprot} tools are not deployed sufficiently~\cite{ceccato-new-one,BSA,GOP,arxan-report}; and expertise is largely missing in software vendors to deploy (third-party) {\softprot} tools~\cite{Gartner-report-online,Irdeto-report1,Mandiant}. 
Moreover, we lack standardization. 
The \nist SP800-39 IT systems risk management standard~\cite{nistSP800-39} or the ISO27k framework for information risk management~\cite{ISO27k}, which are deployed consistently in practice to secure corporate computer networks, have no counterpart or instance in the field of {\softprot}. 
Neither do we have concrete regulations to implement \gdpr compliance in applications.
We can summarize the status of the \softprot domain as an industry with information system business needs involving so-called wicked problems~\cite{hevner2004design}. The foundations and methodologies currently available in the \softprot knowledge base have not met those needs. 

To plug gaps in this knowledge base, most existing \softprot research focuses on piece-wise, bottom-up extensions to its foundations and methodologies by presenting ever more novel \softprot \changed{artifacts} and attack \changed{artifacts} in a \softprot arms race. \changed{That existing offensive and defensive research fits into the information systems \dsr paradigm. Hevner et al.\ define this paradigm as research seeking to ``extend the boundaries of human and organizational capabilities by creating new and innovative artifacts,'' and ``create innovations that define the ideas, practices, technical capabilities, and products through which the analysis, design, implementation, management, and use of information systems can be effectively and efficiently accomplished''~\cite{hevner2004design}}. 

However, to overcome the aforementioned shortcomings and to pave the road towards a standardized risk management approach and automated decision support for \softprot, we are of the opinion that such bottom-up \dsr needs to be complemented with a holistic, top-down design search process in which we study what an end-to-end \softprot risk management approach has to cover and what parts can and should ideally be automated. Our own research hence includes both the bottom-up and the top-down approach in the search for answers to the following research questions (RQs): 
\begin{itemize}\itemsep 0pt 
    \item \textbf{RQ1}: 
    \changed{Can automated decision support tools assist experts with the deployment of {\softprot}s and the use of \softprot tools? Can they also assist non-experts?}
    \item \textbf{RQ2}: To adopt a standardized risk management approach in the domain of SP, which
constructs, models, and methods does the adopted approach need to entail, and which ones thereof should ideally be automated? 
    \item \textbf{RQ3}: Which parts of such an approach can already be automated using decision support tools that instantiate the identified constructs, models, and methods?
\end{itemize}

RQ1 is formulated rather broadly, as usability covers many different aspects such as efficacy, efficiency, and user-friendliness. Later in the paper this RQ will be refined.
For answering it, we developed a \poc decision support tool for \softprot bottom-up, based on concrete requirements and needs from industrial partners of a European research project. Towards RQ2, we explored top-down how a standardized risk management approach can benefit the domain of \softprot. With the birds-eye view of such an approach, we identified existing work to build on and aspects that need more research and/or collaboration in the community. To ensure the relevance of our proposed design, we build on our experience in our academic \softprot research and past collaborations with the industry. That experience allows us to formulate the domain-specific requirements, consider the relevant industrial \sdlc requirements and practices, and position existing domain-specific knowledge in the design. Finally, for formulating a \changed{partial, lower bound} answer to RQ3 \changed{we identified which artifacts from our answer to RQ2 are already instantiated and automated in our \poc tool.}


This paper reports our research findings and presents our answers to the RQs with the following contributions. 
First, we provide a rationale for adopting and standardizing risk management processes for {\softprot}. We discuss several observations on the failing {\softprot} market and we analyse why existing standards are not applicable as is for {\softprot}. 
Where useful, we also highlight differences between \softprot and other security fields such as cryptography, network security, and software security.

Secondly, we discuss in depth how to adopt the \nist risk management approach. We identify which artifacts in the forms of constructs, models, methods, and instantiations, i.e., (semi-)automated tools, we consider necessary and feasible to introduce and deploy the \nist risk management approach for \softprot. For all the required processes, we highlight (i) the current status; (ii) \softprot-specific concepts/artifacts to be covered; (iii) what existing parts can be borrowed from other fields; (iv) open questions and challenges that require further research; (v) needs for the research community and industry to come together to define standards; and (vi) relevant aspects towards formalizing and automating the processes.  

Finally, we demonstrate that several aspects can already be formalized and automated by presenting a \poc decision support system that automates some of the major risk management activities.
Even if not completely automated, this demonstrates that the more abstract constructs, models, and methods we discuss can indeed be instantiated concretely. This \poc  provides a starting point for protecting applications and building a more advanced system that follows all the methodological aspects of a \nist 800-compliant standard with industrial-grade maturity. 
The first results obtained with the tool have been validated mostly positively by industry experts on Android mobile app case studies of real-world complexity and are presented according to the Framework for Evaluation of Design Science~\cite{FEDS-evaluation}. 


%% file: approach.tex
\newpage
\section{Research Approach}
\label{sec:approach}



\begin{figure}
\centering
\includegraphics[width=12cm]{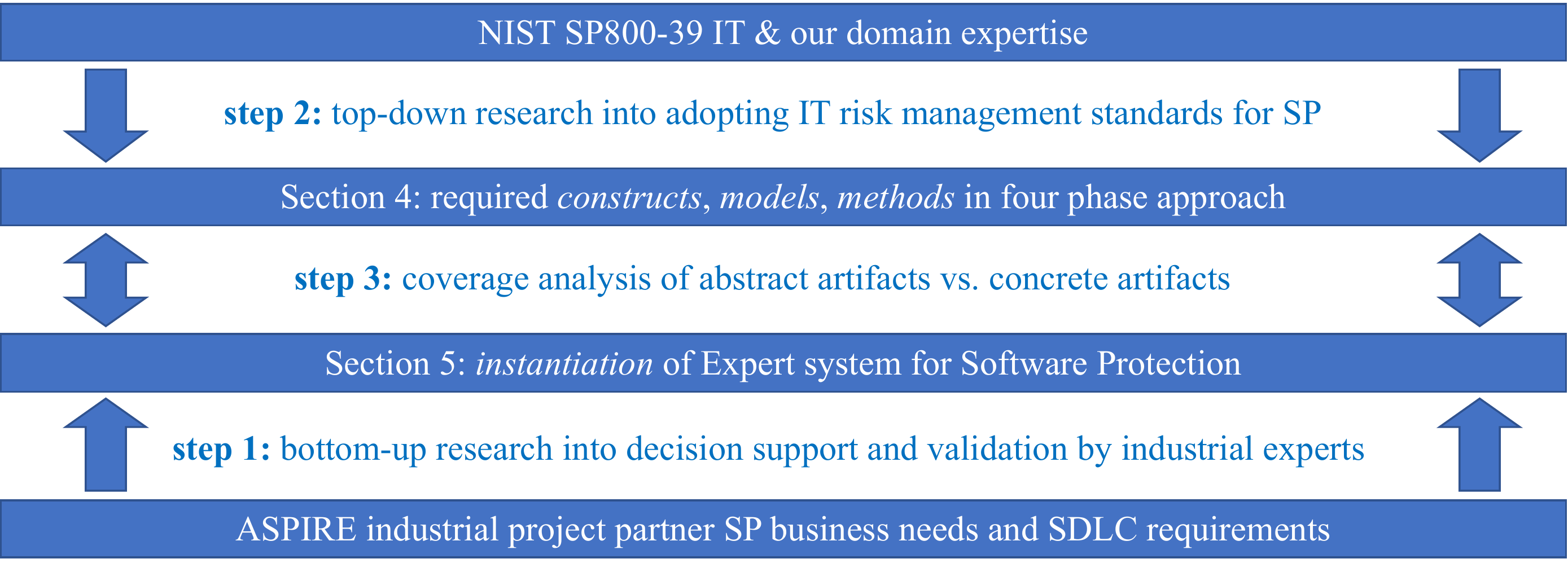}
\caption{Three research steps leading to the results presented in this paper.}
\label{fig:approach}
\end{figure}

\changed{Beyond their quotes in the previous section, Hevner et al.\ describe \dsr as ``achieving knowledge and understanding of a problem domain and its solution by the building and application of designed artifacts''~\cite{hevner2004design}. For doing so in the \softprot risk management problem domain, in particular for answering the three RQs, we collected, structured, designed, built, and applied a large set of related artifacts in the three research steps shown in Figure~\ref{fig:approach}.}

First, in the collaborative European ASPIRE research project, we designed, developed, and evaluated a largely automated \poc called \esp. This bottom-up research was driven by the industrial partners' business needs and \sdlc requirements. Section~\ref{sec:workflow} presents the \poc, of which Section~\ref{sec:evaluation} presents the evaluation.

Second, we studied the adoption of a standardized IT risk management approach, the NIST SP800-39 standard, in the domain of \softprot. This was driven by our observations of the state of the domain of \softprot as discussed in Section~\ref{sec:background} and the motivation presented in Section~\ref{sec:motivation}. The result of this study is the approach presented in Section~\ref{sec:requirements}. 

Third, we analyzed which of the constructs, models, and methods required in the adopted approach are actually covered by the automated tool support in the \esp. The result is a mapping between the artifacts introduced in Section~\ref{sec:requirements} and those discussed in Section~\ref{sec:workflow}.

We now discuss our approach in these steps in more detail.

\subsection{Step 1: Bottom-up Development and Evaluation of Proof-of-Concept Decision Support Tools}
\label{sec:step1:bottom-up}
Our research into \softprot decision support intensified in the 2013--2016 European ASPIRE FP7 research project\footnote{ \url{https://aspire-fp7.eu/}} in which we collaborated with three \softprot companies: Nagravision with a focus on DRM, Gemalto (now Thales) with a focus on software trusted execution environments and SafeNet (now Thales) with a focus on software license management. 
The project researched a layered \softprot toolchain for mobile apps and corresponding (semi-)automated decision support methods and tools. The companies identified the lack of such automated support tools as a critical, foundational gap in the \softprot knowledge base that hampered the effective and efficient deployment of \softprot in practice. In the traditions of \dsr, we endeavoured to close this gap by researching the design and development of novel artifacts, including proof-of-concept tools.
The companies and their technical and commercial \softprot business needs drove the project's requirements analysis and scope determination, as well as the considered attack model. In technical meetings, we engaged with their stakeholders and experts in \softprot, including software developers, \softprot tool developers and users, security architects, and penetration testers. We engaged with higher management in advisory board meetings. The insights obtained there drove the development of decision support techniques in a bottom-up fashion during the ASPIRE project, i.e., starting from concrete \softprot problems and business requirements and solutions, as well as existing, mostly informally described best practices\footnote{Unfortunately, many documents that formalized and structured those insights in the ASPIRE project are confidential.}. As we were, to the best of our knowledge, the first project to research largely automated end-to-end decision support tools, conforming to existing standards was at that time not at all a requirement or concern. 

Through the \softprot tool flow and decision support developed in the project, we provide an answer to RQ1, demonstrating that automated decision support may be within reach. That evidence in the form of artifacts and their evaluation is presented later in this paper. 

We performed and present our evaluation using the \feds~\cite{FEDS-evaluation}, the taxonomy of evaluation methods for IS artifacts by Prat\etal~\cite{Prat-taxonomy}, and the evaluation criteria and terminology by Cleven\etal~\cite{DSRA-evaluation-alternatives}. 

Our \emph{ex-post} evaluation from an \emph{engineering perspective} focused mainly on the \emph{human risk and effectiveness} strategy, as the aim was to determine whether the artifacts consisting of our \poc decision support tool and all the data it generates are accepted by the involved users and whether it benefits their work. We focused on properties compliant with the ISO/IEC 9126-1:2001 criteria\footnote{\url{https://www.iso.org/standard/22749.html}}, namely usability, efficiency, correctness, and comprehensibility and acceptability by the users.
\changed{As Section~\ref{sec:design_qualitative_eval} will describe in much more detail,} we organized the evaluation in multiple steps to gather early feedback and gradually involve more external experts.
Initially, a qualitative assessment of the automatic decision support prototype (and of the artifacts it used) was performed with \changed{three} industrial experts working on the ASPIRE project. The objective was to improve the early versions and components and to release the final prototype. 
A second qualitative assessment was performed on the final \poc, for which we involved \changed{two} additional industrial experts that had not participated in the development.
Moreover, they evaluated the performance of the algorithms and techniques used in the PoC on both the reference use cases and artificial applications, with measurements and with a complexity analysis.
In addition, we performed a \emph{purely technical artifact} assessment to verify that the tool provides solutions in a useful time with reasonable use of resources. 

\subsection{Step 2: Top-Down Adoption of a Standardized IT Risk Management Approach}
After the ASPIRE project had formally finished, we continued our collaboration and gradually developed our vision that the best way to approach decision support is from the perspective of information risk management approaches. This vision first manifested itself in the July 2019 Ph.D. thesis of Leonardo Regano~\cite{ReganoPhd} that presents the components of the \esp and that is structured according to the phases of information risk management standards. 

We reached out to other researchers and practitioners in the \softprot domain to gather their opinions and insights, as well as doubts on decision support for \softprot. This happened in informal discussions but also in structured ones, including the August 2019 Dagstuhl seminar on Software Protection Decision Support and Evaluation Methodologies~\cite{Dagstuhl}, of which B.\ De Sutter was the main organizer. In this one-week seminar, the three senior authors of this paper engaged again with a range of experts, including, amongst others, \softprot researchers, security economists, reverse engineering practitioners, software analysis experts, and commercial \softprot developers. 
During the seminar, the need for standardization came to the forefront, if not formalized, then at least in terms of best practices and guidelines for conducting research into SP and evaluating the strength of proposed SPs and attacks thereon. 

Following that seminar, we invested in a top-down approach, studying the adoption of existing information and system security standards in the domain of \softprot. We investigated how the generic concepts that make up these standards are specialized and adopted in specific security domains such as network security. We then extensively brainstormed about how they can also be specialized and adopted in the domain of \softprot. For example, the \nist SP800-39 IT systems risk management standard~\cite{nistSP800-39} prescribes a top-level method consisting of four generic risk management phases, each corresponding to their own, conceptually formulated, abstract method. We studied the domain-specific organizational problems that need to be solved in the different phases, and which more concrete domain-specific concepts those phases need to encompass for \softprot. 

We started this research by collecting our combined insights, then structuring them, and then iteratively coming to the text of Section~\ref{sec:requirements} that, in essence, presents a top-level \softprot risk assessment method and the necessary artifacts for using that method, thus providing our answer to RQ2. We want to thank the reviewers of earlier versions of this text for their valuable insights that helped us produce the final result. 

While this research step was rooted in our previous experience with \softprot, we tried to perform this study as independently as possible from the \poc results of the ASPIRE project. This shows, amongst others, in the fact that in Section~\ref{sec:requirements}, we put forward about 40\% more concepts to be included in the proposed standard approach adoption than are covered in the \poc results we present in Section~\ref{sec:workflow}. As a concrete example, we discuss the organizational problem of \softprot tool vendors and their customers not giving each other white-box access because they do not trust each other in Section~\ref{sec:framing_sdlc}. That problem was out of scope in the ASPIRE project and is hence not tackled by the presented \poc tools.

\subsection{Step 3: Coverage Analysis of the Adopted Approach in the \dsr Framework}

An important consideration in our study in step~2 was the need for automation, as reflected in the last part of RQ2 and in RQ3. In later sections, we argue in more detail why we consider automation of many of the adopted and specialized methods beneficial, if not crucial.

Our answer to RQ3 is not based on theoretical analysis and abstract reasoning but on tangible evidence, i.e., the existence of the concrete artifacts that form the \poc developed in step 1. To provide this answer to RQ3, we organized numerous internal discussions in which we analyzed which of the concepts from the different phases of the proposed approach are instantiated by the automated components of our \poc. 

To do this more methodologically, we adopted the \dsr framework by Hevner et al.~\cite{hevner2004design}. First, we rephrased the adopted approach such that all essential concepts of the approach's four phases are clearly identified as either constructs (vocabulary to define and communicate concrete \softprot cases), models (abstractions and representations to aid case understanding and to link case features and solution components to enable exploration), and methods (algorithms, practices, and processes, as well as guidance on how to tackle concrete cases). These are the three abstract types of artifacts that Hevner et al.\ identify as foundational elements of an information systems knowledge base, in this case the \softprot knowledge base. 

Next, we identified which of these abstract artifacts are instantiated by means of components of the \poc \esp. Such implementations are called instantiation artifacts by Hevner et al. They form the fourth type of foundational element in a knowledge base. 

The mapping from more abstracts \dsr artifacts onto concrete \poc instantiation artifacts is documented in this paper by means of recurring tags. The tags are introduced in Section~\ref{sec:requirements} when the constructs, models, and method artifacts are first introduced, and they recur in Section~\ref{sec:workflow} where the corresponding instantiations are discussed.

%% file: rationale.tex
\section{Background on Standardization and the State of Software Protection}
\label{sec:background}


We first discuss some risk management standards and how they have been adopted in other security domains, such as network security, and the healthy market for products and services that exists there as a result. We then contrast this with the lack of such a market and standards for {\softprot}.

\subsection{Standardized Risk Management Approaches}
\label{sec:standardizedRisk}
Protecting software can be seen as a risk management process, a customary activity in various industries such as finance, pharmaceutics, infrastructure, and \itech. 
The \nist has proposed an \itech systems risk management standard that identifies four main phases \cite{nistSP800-39}:
\begin{enumerate}\itemsep 0pt
\item \emph{risk framing}: to establish the scenario in which the risk must be managed;
\item \emph{risk assessment}: to identify threats against the system assets, vulnerabilities of the system, the harm that may occur if those are exploited, and the likelihood thereof;
\item  \emph{risk mitigation}: to determine and implement appropriate actions to mitigate the risks;
\item \emph{risk monitoring}: to verify that the implemented actions effectively mitigate the risks.
\end{enumerate}

The ISO27k framework also focuses on information risk management in three phases~\cite{ISO27k}:
\begin{enumerate}\itemsep 0pt
\item \emph{identify risk} to identify the main threats and vulnerabilities that loom over assets;
\item \emph{evaluate risk} to estimate the impact of the consequences of the risks; 
\item \emph{treat risk} to mitigate the risks that can be neither accepted nor avoided.
\end{enumerate}
ISO27k adds an explicit operational phase for handling changes that happen in the framed scenario.

Those approaches have been consistently applied in practice for securing corporate networks.
Regulations stimulate companies to analyse the risks against their \itech systems. For instance, the \gdpr explicitly requires a risk analysis of all private data handling. 
Companies invest in compliance with the ISO27k family to obtain market access.
Consequently, risk analysis of networks has developed a common vocabulary, and a company's tasks have been properly identified and often standardized, so offerings from consultancy firms can be compared easily. 
There is a business market related to this task, best practices, and big consultant firms have risk analysis of corporate networks in their catalogs~\cite{Gartner-report-riskanalysis}.

In the domain of software security, several frameworks for risk analysis and decision support exist that mainly focus on Software Vulnerability Management~\cite{nistir8011} and Enterprise Patch Management~\cite{nistSP800-40}. 
Other frameworks focus on quality assurance best practices and benchmarking, including the OWASP Software Assurance Maturity Model (SAMM)~\cite{owaspsamm}, the OWASP Application Security Verification Standard (ASVS)~\cite{owaspasvs}, and the Building Security in Maturity Model (BSIMM)~\cite{bsimm}. 
These address problems of software security and are not applicable to {\softprot}.

NIST SP800-53~\cite{nistSP800-53} extends beyond software security and provides a comprehensive and flexible catalog of privacy and security controls for systems and organizations as part of their organizational risk mitigation strategy, for which they build on NIST SP800-39~\cite{nistSP800-39}. It targets whole IT infrastructures, including hardware and software. Regarding software, it advises to "Employ anti-tamper technologies, tools, and techniques throughout the system
development life cycle" in its SR-9 Supply Chain Risk Management family of controls. Obfuscation is mentioned only as an option to strengthen the tamper protection, not to protect the original software. The document does not discuss how to deploy these protections, or how to select the ones to deploy. NIST SP800-53 is hence not applicable to {\softprot}. For much of the remainder of this paper, we will actually discuss what a {\softprot} counterpart of NIST SP800-53 needs to entail. 



\subsection{The State of \mate Software Protection}


Compared to network security and software security, \softprot has years of delay. 
For setting the scope, 
Table~\ref{tab:protection_examples} lists a number of well-known {\softprot}s. Out-of-scope are mitigations to prevent the exploitation of vulnerabilities, such as \aslr, compartmentalization techniques, or safe programming language features in, e.g., Rust. 
In the \mate attack model, attackers have full control over the devices on which they attack the software. They can disable security features of the operating system and the run-time environment, such as \aslr, which therefore cannot be trusted. 
For that reason, {\softprot} centers around protections embedded in the software itself, rather than relying on the security provided by the run-time environment.

\begin{table}[t]
    \centering
    {
    \setlength{\tabcolsep}{0.3em}
    \small
    \begin{tabular}{lp{10.2cm}}
        \toprule
        protection type & explanation\\ 
        \midrule
            anti-debugging             & Techniques to detect or prevent the attachment of an attacker's debugger~\cite{circulardebugging}. \\
            branch functions           & Indirect, computed jumps replace direct control transfers to prevent reconstruction of control flow graphs~\cite{linn2003branchFunctions}.\\
            call stack checks          & Checks if functions are called from allowed callers to block out-of-context calls. \\
            code mobility              & Code is lifted from the binary to prevent static analysis. At run time, the code is downloaded into the running app from a server~\cite{codeMobility}. \\
            code virtualization        & Code in the native instruction set is replaced by bytecode and an injected interpreter interprets that bytecode, of which the format is diversified~\cite{Anckaert2006}.\\
            control flow flattening    & A structured control flow graph graph is replaced by a dispatcher that transfers control to any of the original nodes based on data. This makes it harder to comprehend the original flow of control and the code~\cite{wangFlatteningTechReport}.\\
            data obfuscation           & Transformations that alter data values and structures to hide the original ones.\\
            opaque predicates          & Logic that evaluates to true/false based on invariants known at protection time but that are hard to discover by an attacker~\cite{collbergOpake}. This enables inserting bogus control flow to hinder code comprehension and precise analysis~\cite{JENS,Jens2}.\\
            remote attestation         & Techniques in which a remote server sends attestation requests to a running program. If the program fails to deliver valid proof of integrity, it is considered to be tampered with, and an appropriate reaction can be triggered~\cite{viticchie2016reactive}. \\
            white-box crypto     & Implementations of cryptographic primitives such that even white-box access to the run-time program state does not reveal the used keys~\cite{wyseur2011white}.\\
        \bottomrule
    \end{tabular}
    }
    \caption{A number of software protections.}
    \label{tab:protection_examples}
\end{table}


The market of \changed{such} {\softprot} is neither open nor accessible to companies with a small budget.
In 2017 Gartner projected that 30\% of enterprises would have used \changed{\softprot} to protect at least one of their mobile, IoT, and JavaScript critical applications in 2020~\cite{Gartner}.
However, two years later Arxan reported that 97\% (and 100\% of financial institutions) of the top 100 mobile apps are easy to decompile as they lack binary code protection or implement weak protection~\cite{arxan-report}. 
A study confirms the absence of both anti-debugging and anti-tampering protections for 59\% of about 38k Play Store apps. The study highlights that weak Java-based methods are employed in 99\% of the {\softprot} uses~\cite{ceccato-new-one}. Repackaging benign apps to obtain malicious apps~\cite{Khanmohammadi2019repacked,Zhou2012repacked} is easy because of the intrinsically weak app packaging process but also because used anti-repackaging protections are currently weak~\cite{merlo2021repackage}.
Furthermore, it is estimated that 37\% of installed software is not licensed, for a total amount of losses estimated at \$46.3B in 2015--2017~\cite{BSA}.
Consequently, the \softprot market, which accounted for \$365.4M dollars in 2018, is expected to grow fast~\cite{frost}.

Cybersecurity competences are lacking~\cite{Gartner-report-online}. \softprot is no exception.
Few companies have internal {\softprot} teams: only 7\% of respondents stated their organization has all it needs to tackle cybersecurity challenges; 46\% stated they need additional expertise/skills to address all aspects of cybersecurity~\cite{Irdeto-report1}. Meanwhile, many organizations lack competent staff, budget, or resources~\cite{Mandiant}.

When the value of assets justifies it, developers resort to paying third parties to protect their software. 
The price is typically high, involving licenses to tools and often access to expert consultants. Moreover, the services and the strength of the obtained \softprot are covered by a cloak of opaqueness, with \sto omnipresent\footnote{
Abandoning \sto implies that transparency is given about the \softprot process, the design and implementation of all \softprot tools being used, including the supported {\softprot}s and decision support tools. It does not at all imply that \softprot users need to be transparent about the applications they protect. Indeed, the very objective of using \softprot to hamper MATE attacks on assets with confidentiality requirements is to keep those assets obscured. This is to be achieved by keeping the unprotected code secret, and by keeping the used tool configuration secret, not by hiding the used tools or evaluations of their effectiveness.  
}.
For example, whereas early white-box cryptography schemes were peer reviewed~\cite{AESwhite,chow2002white} and then broken~\cite{AESbroken,DESbroken}, we could not find peer-reviewed analyses of schemes currently marketed by big vendors. 
Moreover, most vendors' licenses forbid the publication of reverse engineering and pen testing reports on their products. They do not share their internal procedures, tools, or reports with academics. 
%

We deduce that many companies do not understand the risk and therefore do not feel the need for deploying {\softprot}, or they do not have the internal competences and knowledge to do so properly, or they lack the money to pay third-party providers. In short, there exists no widely accessible, functional, transparent, open {\softprot} market. 
At some of the big {\softprot} vendors that are also active in other security fields, risk analysis and mitigation is most certainly the principle that drives their experts and that is encoded in policies. Yet no methodology is publicly available for applying a risk analysis process when deciding how to protect software.
Needless to say, no standard process guarantees the proper selection and application of available {\softprot}s given a case at hand.

\section{Motivation and Challenges for Standardization, Formalization, and Automation}
\label{sec:motivation}

This section first motivates why we strive for standardization. Next, it argues why formalization and automation are (equally) important. The section concludes with a discussion of some challenges towards these objectives, thus complementing the background provided above.

\subsection{Motivation for Standardization}

Standardization efforts aim at ``striking a balance between users' requirements, the technological possibilities and associated costs of producers, and constraints imposed by the government for the benefit of society in general''~\cite{standardization}. The benefits come from the positive impact of standards on quality/reliability, information standards, compatibility/interoperability, and variety reduction~\cite{standardization}. In line with those benefits, a standardized, methodological approach to \mate risk analysis could have a plethora of benefits. This section speculates on this potential.


First, it could force stakeholders to follow a more rigorous approach to \softprot. Risk framing forces analysts to define workflows, processes, methods, and formulas to evaluate risks and the impact of mitigations. In network security, a structured risk analysis has limited the impact of subjective judgments by suggesting the use of collegial decisions involving more roles \cite{nistSP800-39}.
A more rigorous approach for \softprot could similarly increase the transparency of all phases, guaranteeing a more reliable estimation of the reached \softprot level and of the quality delivered by third parties.
In turn, we expect less reliance on \sto. 
Simply adopting the OWASP Security Design Principles forces security specialists to avoid \sto, which is also considered a weakness in MITRE CWE 656~\cite{CWE-656}.

A standard could induce the community to use well-defined terminology and to agree on the meaning of each {term}, as happened after  NIST SP 800~\cite{nistSP800-39}.
Building common ground and well-defined playing rules would also benefit the {\softprot} market by creating a more open and transparent ecosystem where services can be compared as normal products, thus bridging the gap with the network security market in which products are evaluated by third parties using standardized methods such as the Gartner Magic Quadrant for Network Firewalls~\cite{magicquadrant}.
Hence, we expect the rise of consultancy firms that can independently evaluate \softprot effectiveness.
We also expect a price reduction, as highlighted in a study~\cite{frost}. 
With a lower entry price and the definition of entry-level protection services, more companies can then afford professional \softprot services, with benefits for all the stakeholders.

When \softprot becomes standardized and more clearly defined, it could also create a market for decision support products that automate risk management. This could in turn lead to cost savings and to more accessible and more effective \softprot. 

The availability of standards increases awareness, as reported by an EU agency one year after adopting the GDPR \cite{gdpr1yafter}.
The mere existence of a standard would initially inform people about the need for \softprot.
Compliance would then force all parties to obtain in-depth knowledge, and the standards and related best practices would eventually be incorporated into educational programs. 

The work towards standards could also impact research. 
It could initially stimulate the community to focus on identifying and plugging existing gaps and, later on, create new or more effective, validated {\softprot}s to be integrated into a standard framework. 
The interest in the field and the impact of research results would then likely attract more researchers to the {\softprot} field, which is now marginal in the software engineering community. 
We have found analogies with the impact of the ISO/SAE~21434 standard for cybersecurity engineering of road vehicles \cite{10.1007/978-3-030-55583-2_9}.  
Years before its adoption, car manufacturers anticipated effort and funded research 
to cope with the demanding standard. The investments in automotive cybersecurity will grow from \$4.9B in 2020 to \$9.7B in 2030, with a market size expected to grow from \$238B in 2020
to \$469B in 2030~\cite{mckinsey-automotive}. Parts of this increase and of the focal shift towards cybersecurity might not be caused directly by the ISO/SAE~21434 standardization. However, we are convinced that the planned standardization was a major contributing factor in the past years, given that compliance with the standard as part of UN R155 has already become mandatory in Europe, Japan, and Korea since July 2022. The anticipation of the standard can also be observed in guidelines published long before its finalization, such as in the "ENISA good practices for security of Smart Cars" published in November 2019 with contributions of major carmakers~\cite{ENISA}.



Increased attention by research institutions and academia usually translates into better education opportunities, possibly with dedicated curricula, which usually pair well with the career opportunities created by a more open market.
Ultimately this could help companies employ skilled people and support a freer job market to compensate at least partially for the lack of \softprot experts.


In the end, the benefit would extend to the whole society, as having better-protected software reduces the global exposure of citizens to risks and, we hope, would make \mate attacks a less lucrative field, or at least reduce its growth.

\subsection{Motivation for Formalization and Automation}
\label{sec:motivation_formalization_automation}
A standardized, methodological risk management approach is not necessarily formalized or automated. We argue, however, that formalization and automation are by and large required. The main reason is the need for precision, i.e., the repeatability or reproducibility of obtained results. 

In the security field, including {\softprot}, we want to avoid a scenario in which different experts that deploy the same risk management approach on the same software under the same conditions would come up with different sets of identified threats and different sets of supposedly good combinations of protections. One of the more important reasons to stay clear of such a scenario is that it would complicate the validation and enforcement of compliance. 

Cognitive psychology research has shown, however, that humans are incorrigibly inconsistent in making summary judgments based on complex information~\cite{thinking,rational}. Hence they provide different answers when asked to evaluate the same information multiple times. Experts also suffer from this. Their judgments hence lack precision in environments that are not sufficiently regular to be predictable~\cite{conditions,thinking}. Those environments are also known as low-validity environments. Determining the major \mate attack threats on a given piece of software given the source code, the formulated security requirements, the domain knowledge, etc., as well as selecting appropriate combinations of {\softprot}s come down to making predictions in such an environment. One of the reasons is that there are many parameters one cannot think of in advance, such as the configurations with which the final software will be deployed on-site. Psychology research has also shown that the precision of expert judgment improves when there exists backup in the form of formulas and algorithms to complement, guide, or replace otherwise imprecise human cognitive processes~\cite{robust}. We hence put forward formalization and automation as important objectives for \mate risk management. 

We are not the first ones to do so. For example, in their survey on architectural threat analysis, Tuma et al.\ analyse whether the surveyed methods are supported by formal frameworks and by (semi-)automated tools because of their impact on precision~\cite{architectural}. They also differentiate between template-based approaches and example-based ones, as the former yield higher precision. 
Similarly, we put forward that using an unambiguous vocabulary with clear definitions will benefit the precision of \mate risk management.

Economic arguments further support our claim that automation cannot be separated from the aim of adopting a risk analysis process for \softprot. Manual \softprot decision making requires expertise, effort, and hence time. As we discussed, there are not enough experts to protect all software that can benefit from rigorous \softprot. Even if enough experts were available to put in the necessary manual effort, they would remain costly, keeping good \softprot out of reach for SMEs. 

Scaling up the number of experts to meet all demands without automating parts of the processes is not realistic. Every time a new version of an application is issued (e.g., because of regular updates or a bug), it needs to be protected. Part of the work on previous versions can probably be reused, but typically the {\softprot}s at least need to be diversified.

Additionally, \softprot firms may have to protect many versions, such as ports of the same software to different platforms, including laptops or mobiles with limited computational power. If maintaining the application's usability is at risk on some platforms because of the \softprot overhead, developers may decide to limit the features on those platforms. As an example, media players with DRM will only access low-quality versions of media if the platform does not allow full protection.

Moreover, even if human experts were available, their latency would still be problematic. Software vendors face time-to-market pressure. For that reason alone, automated tool support that can cut the time and effort required to protect applications is beneficial.

\subsection{Challenges towards Standardization, Formalization, and Automation}
\label{sec:challenges}

\input{challenges-aldo}

%% file: challenges-aldo.tex


Despite the many benefits a standardized, formalized, and automated approach would bring, such an approach is a long way off, and adopting a \nist-style risk management faces several challenges.

%
%
\changed{
A first challenge relates to the definition of asset categories and their relation with security properties. These are lacking today, which is problematic for the framing of risks. %
{\softprot} relies on the \mate attacker model that has never been defined clearly. The abilities of \mate attackers are unclear, not in the least because of the complexity of modelling human code comprehension and software tampering capabilities.

A second challenge is the definition of threat and risk assessment models that allow enough precision and objectivity. Estimating the feasibility of \mate attacks requires a white-box analysis of the assets and of the entire application.
The complexity of mounting static, dynamic, symbolic, and concolic attacks heavily depends on the structure and artifacts of the software, such as the occurrence of all kinds of patterns or observable invariants. 

Thirdly, moving towards a more precise categorization of protections and risk in the \mate scenario is another challenge that needs to be overcome for the risk mitigation phase. In practice, \softprot provides only fuzzy forms of protection. {\softprot}s have only been categorized coarsely (e.g., obfuscation vs.\ anti-tampering). In general, it is not clear what security level they offer where, and there yet exists no well-defined set of categories of security controls to mitigate \mate risks. This contrasts with, e.g., the field of cryptography, in which algorithms are characterized in terms of well-defined properties such as ciphertext indistinguishability or second pre-image resistance~\cite{intro_crypto}. Also in network security, it is clear what firewalls and VPNs do and how to use them to mitigate network security risks. There are accepted measures and guidelines to estimate the effectiveness of categories of network security mitigations and in some cases categorization of tools and vendors that help in estimating their efficacy~\cite{ISO27004}. The \mate domain lacks such well-definedness. 

Fourthly, today it remains a huge challenge to simply measure or estimate the efficacy of {\softprot}s. This is obviously necessary to assess the residual risks of deployed {\softprot}s. However, no metrics are currently available to quantify \softprot efficacy. Potency, resilience, and stealth are commonly accepted criteria~\cite{collberg1997taxonomy}, but no standardized metrics are available for measuring them. Complexity metrics originating from the field of software engineering have been proposed~\cite{D4.06}, and ad-hoc metrics are used in academic papers~\cite{JENS,linn2003branchFunctions}. However, none have been empirically validated for use in {\softprot}, and practitioners most often do not see the metrics used in academic papers as reliable proxies of real-world potency, resilience, or stealth. Using those metrics is hence not yet considered a viable replacement for human expertise and manual pen testing. In many cases, there are no hard proofs that {\softprot}s are effective in delaying attackers. Rather than encouraging checks by external parties, {\softprot} vendors often contractually prevent the analysis of protected code, instead relying on \sto. As a result, there is neither an objective nor a measurable assurance of protection, nor an objective evaluation of the companies' work.
In academic {research}, the situation is not much better. For example, the seminal obfuscation versus analysis survey from Schrittwieser et al.\ never refers to a risk analysis framework~\cite{schrittwieser2016}. Their results, although widely acknowledged, are hence not readily usable in a decision process.

The aforementioned challenges are particularly hard because in \softprot, determining the boundaries between assets and protections is no easy task. \softprot{}s are often processes that transform assets to hinder analysis and comprehension of their logic~\cite{schrittwieser2016}.
For instance, most forms of obfuscation transform code fragments. Since {\softprot}s need to be layered for stronger and mutual protection and to exploit synergies, obfuscation can transform code that results from previous transformations, such as code guards injected for anti-tampering purposes. Some obfuscations even aim for eliminating recognizable boundaries between different components~\cite{JENS}, and others aim for re-using application code for obfuscations~\cite{Jens2}. 
As a result, the code of multiple {\softprot}s and of the assets they protect becomes highly interwoven. We hence need to talk of protected assets, certainly not of separated protection and asset entities. 

Furthermore, software internals must be known to the tools. This includes the types of instructions, structure and semantics of the code, and the presence of any artifacts that might benefit attackers. This information is needed to decide whether some (layered) {\softprot} can be effective or not and to tune its parameters.
In addition, it is generally accepted that in order to deploy {\softprot}s effectively, an application's architecture needs to be designed with the protection of the sensitive assets in mind.
If it is not designed well, {\softprot}s will only provide superficial mitigation.  For theoretical definitions of \softprot, such as virtual black-box obfuscation, Barak already proved this for contrived programs~\cite{barakImpossibility}, but this statement also holds for real-world software and practical \softprot. Examples of design problems that are hard, if not impossible, to fix with {\softprot}s are bad external or internal APIs, missing authorization, and improper or missing crypto key ladders to protect various assets. Such ladders require complex key management, key storage, and crypto functionality, which are easy to get wrong for non-experts. Risk assessment methods must hence recognize software whose design prevents proper protection and report that risks cannot be reduced to the desired level solely with {\softprot}s.
This again stresses that \mate risk analysis requires insights into software internals to identify weaknesses that may turn into vulnerabilities that cannot be protected with \softprot.

\softprot thus poses challenges that, to the best of our knowledge, have not emerged and have thus not been addressed when standardizing other fields such as the network security field in which de-facto standards are used and referred to. These challenges impact the standardization of risk management and, in particular, the definition of objective criteria for assessing the mitigations.

In conclusion, despite their obvious appeal, risk management standardization and a functioning open market as they exist in other areas of ICT security are in our opinion missing in {\softprot} not only because the community is late in developing them, but also because managing the risks in \softprot is really challenging. 
}

%% file: requirements.tex
\section{Adopting a Standard Towards Proper Risk Management}
\label{sec:requirements}

\begin{figure}[t]
\begin{center}
\includegraphics[width=11cm]{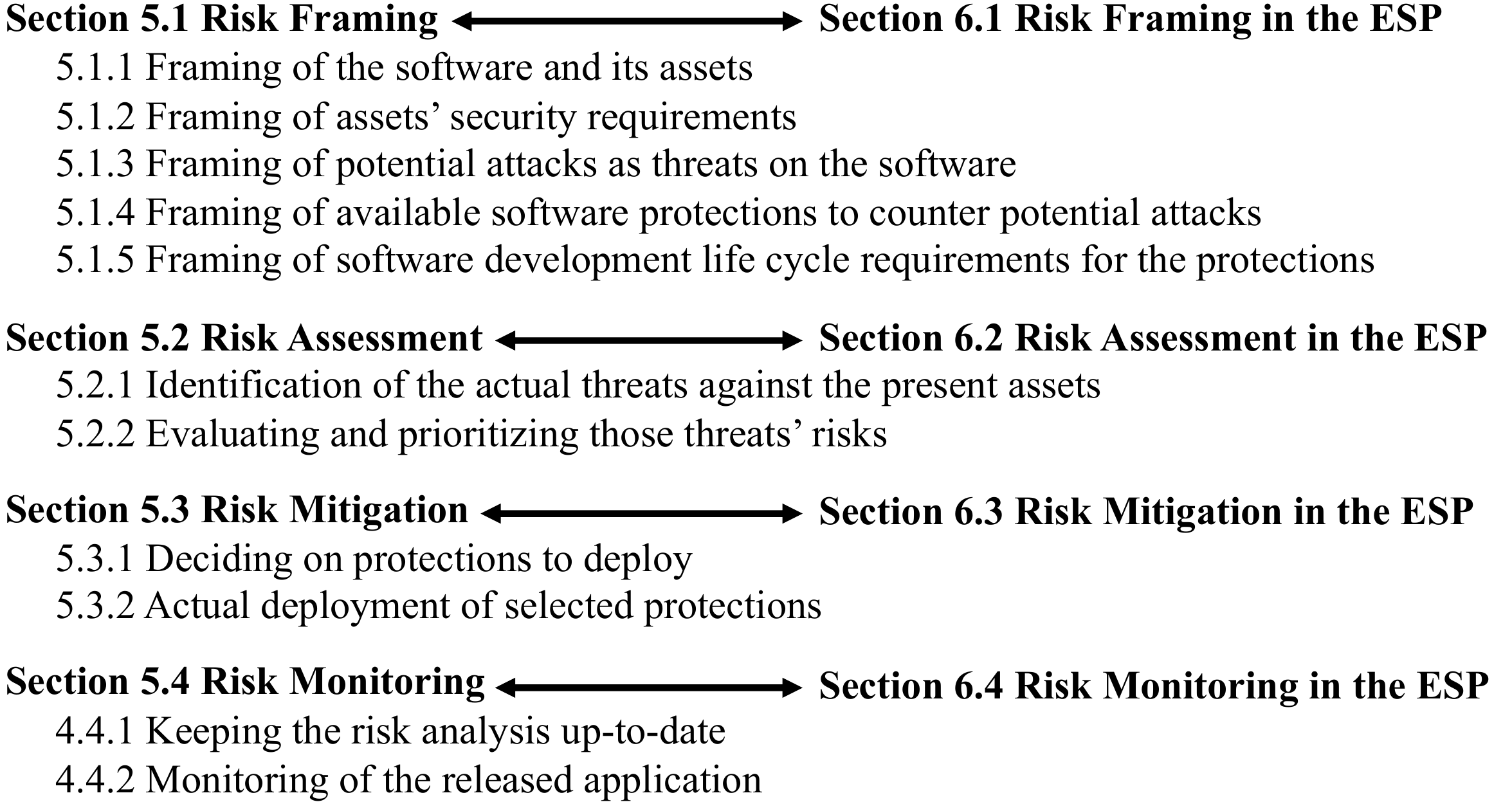}
   \caption{Four phases of the proposed risk management approach with reference to the corresponding sections in the presentation of the approach in Section~\ref{sec:requirements}) and in the presentation of the \poc implementation in Section~\ref{sec:workflow}.}
   \label{fig:flow}
\end{center}

\end{figure}

This section provides an answer to RQ2 by discussing what the four phases of the \nist IT systems risk management standard would entail as applied to {\softprot}, i.e., what tasks need to be done in its four phases. Figure~\ref{fig:flow} presents an overview. Note how the tasks flow quite naturally, each task building on the previous ones. 
The discussion of these tasks will cover various recurring aspects, which are highlighted by means of numbered text markings. We introduce the necessary \introconstruct{constructs}\footnote{In most cases, we identify only the abstract top-level constructs, under which more concrete constructs have to be included as well. For example, we will mention the "software protection" construct, without enumerating concrete protections such as opaque predicates, control flow flattening, virtualization, etc.}, \intromodel{models}, and \intromethod{methods/practices}, introducing some useful new terminology along the way.

Tables~\ref{tab:constructs},~\ref{tab:models}, and~\ref{tab:methods} present an overview of the covered abstract artifacts. For those artifacts that have already been implemented in an actual instantiation, the ESP column lists the subsections of Section~\ref{sec:workflow} in which that instantiation will be discussed in more detail. Those instantiations will demonstrate that these artifacts can in fact be implemented in a working system. They will hence demonstrate the feasibility of the covered artifacts, thus also enabling a concrete assessment of their suitability for their intended purpose, as will be discussed in Section~\ref{sec:esp:results}.

\begin{table}[t]
\begin{center}
{\footnotesize
\begin{tabular}{r l c | r l c}
No. & Construct Name & ESP & No. & Construct Name & ESP \\
\hline
\constructlab{1}  & \hyperlink{construct.1}{primary asset}                                      &   \ref{sec:esp:risk_framing}   & 
\constructlab{26} & \hyperlink{construct.26}{protection applicability}                            &   \ref{sec:esp:risk_framing}  \\
\constructlab{2}  & \hyperlink{construct.2}{secondary asset}                                     &  \ref{sec:esp:risk_framing}    & 
\constructlab{27} & \hyperlink{construct.27}{protection composability}  &   \ref{sec:esp:risk_framing},\ref{sec:esp:risk_mitigation} \\
\constructlab{3}  & \hyperlink{construct.3}{attack path}                                         &   \ref{sec:esp:risk_assessment}   & 
\constructlab{28} & \hyperlink{construct.28}{layered protection deployment}  &    \ref{sec:esp:risk_framing},\ref{sec:esp:risk_mitigation} \\
\constructlab{4}  & \hyperlink{construct.4}{attack step}  &   \ref{sec:esp:risk_framing},\ref{sec:esp:risk_mitigation}   & 
\constructlab{29} & \hyperlink{construct.29}{protection synergies}  &   \ref{sec:esp:risk_framing},\ref{sec:esp:risk_assessment}  \\
\constructlab{5}  & \hyperlink{construct.5}{attack pivot}                                        &                & 
\constructlab{30} & \hyperlink{construct.30}{potency}           &   \ref{sec:esp:risk_framing},\ref{sec:esp:risk_mitigation} \\
\constructlab{6}  & \hyperlink{construct.6}{attack time frame}                                   &                & 
\constructlab{31} & \hyperlink{construct.31}{resilience}                                          &   \ref{sec:esp:risk_framing} \\
\constructlab{7}  & \hyperlink{construct.7}{asset renewability}                                  &                & 
\constructlab{32} & \hyperlink{construct.32}{stealth}                                             &  \ref{sec:esp:risk_framing}  \\
\constructlab{8}  & \hyperlink{construct.8}{primary security req.}                                &    \ref{sec:esp:risk_framing}  & 
\constructlab{33} & \hyperlink{construct.33}{overhead/cost constraints}  &   \ref{sec:esp:risk_framing},\ref{sec:esp:risk_mitigation}  \\
\constructlab{9}  & \hyperlink{construct.9}{non-functional security req.}                         &    \ref{sec:esp:risk_framing}  & 
\constructlab{34} & \hyperlink{construct.34}{software development life cycle req.}        &    \ref{sec:esp:risk_framing} \\
\constructlab{10} & \hyperlink{construct.10}{attack identification phase}                         &                & 
\constructlab{35} & \hyperlink{construct.35}{profile information}                                 &     \\
\constructlab{11} & \hyperlink{construct.11}{attack exploitation phase}                           &                & 
\constructlab{36} & \hyperlink{construct.36}{software connectivity}                                   &   \ref{sec:esp:risk_framing}  \\  
\constructlab{12} & \hyperlink{construct.12}{secondary security req.}                             &   \ref{sec:esp:risk_framing}   & 
\constructlab{37} & \hyperlink{construct.37}{software update ability}                                        &     \\   
\constructlab{13} & \hyperlink{construct.13}{functional security req.}                            &     & 
\constructlab{38} & \hyperlink{construct.38}{environment limitations}                             &   \ref{sec:esp:risk_framing}  \\
\constructlab{14} & \hyperlink{construct.14}{assurance security req.}                             &   \ref{sec:esp:risk_framing}  & 
\constructlab{39} & \hyperlink{construct.39}{actual threats}                                      &   \ref{sec:esp:risk_assessment}  \\
\constructlab{15} & \hyperlink{construct.15}{protection policy req.}                              &                & 
\constructlab{40} & \hyperlink{construct.40}{actual risks}                                        &    \ref{sec:esp:risk_assessment} \\
\constructlab{16} & \hyperlink{construct.16}{weaknesses}                                          &                & 
\constructlab{41} & \hyperlink{construct.41}{attack surface}                                      &    \ref{sec:esp:risk_assessment} \\
\constructlab{17} & \hyperlink{construct.17}{attack resources}                                    &   \ref{sec:esp:risk_framing}   & 
\constructlab{42} & \hyperlink{construct.42}{attack vectors}                                      &   \ref{sec:esp:risk_assessment}  \\
\constructlab{18} & \hyperlink{construct.18}{attack capabilities}                                 &   \ref{sec:esp:risk_framing}   & 
\constructlab{43} & \hyperlink{construct.43}{attack paths of least resistance}                    &     \\
\constructlab{19} & \hyperlink{construct.19}{worst-case scenario assumptions}                     &                &
 \constructlab{44} & \hyperlink{construct.44}{analysis tools / toolbox}  & \ref{sec:esp:risk_framing},\ref{sec:esp:risk_assessment}  \\
\constructlab{20} & \hyperlink{construct.20}{attack enabling features}                            &    \ref{sec:esp:risk_framing}  & 
\constructlab{45} & \hyperlink{construct.45}{software features}                   &    \ref{sec:esp:risk_assessment} \\
\constructlab{21} & \hyperlink{construct.21}{attack preventing features}                         &    \ref{sec:esp:risk_framing}  & 
\constructlab{46} & \hyperlink{construct.46}{third-party-provided incomplete analysis}            &     \\
\constructlab{22} & \hyperlink{construct.22}{attack effort determination features}  &    \ref{sec:esp:risk_framing},\ref{sec:esp:risk_assessment}  & 
\constructlab{47} & \hyperlink{construct.47}{residual risks}                                      &   \ref{sec:esp:risk_mitigation}  \\
\constructlab{23} & \hyperlink{construct.23}{attack likelihood of success features}       &   \ref{sec:esp:risk_framing}  & 
\constructlab{48} & \hyperlink{construct.48}{most protective protection solution}      &    \ref{sec:esp:risk_mitigation:optimization} \\
\constructlab{24} & \hyperlink{construct.24}{software protections}                                &    \ref{sec:esp:risk_framing}  & 
\constructlab{49} & \hyperlink{construct.49}{alternative protection targets}  & \ref{sec:esp:risk_mitigation:optimization} \\
\constructlab{25} & \hyperlink{construct.25}{protection strength metrics}  & \ref{sec:esp:risk_framing},\ref{sec:esp:risk_mitigation} & 
\constructlab{50} & \hyperlink{construct.50}{mitigation round}         &     \\  
\end{tabular}
}
\caption{Constructs of the proposed approach, with references to the discussions of their instantiation, if any.}
\label{tab:constructs}
\end{center}
\end{table}

\begin{table}[t]
\begin{center}

{\footnotesize
\begin{tabular}{r l c | r l c}
No. & Model Name & ESP & No. & Model Name & ESP \\
\hline
\modellab{1}  & \hyperlink{model.1}{application and asset model}   &  \ref{sec:esp:risk_framing}  &  
\modellab{4} & \hyperlink{model.4}{attack model} & \ref{sec:esp:risk_framing}     \\
\modellab{2}  & \hyperlink{model.2}{secondary asset attributes model}  &  \ref{sec:esp:risk_framing}     & 
\modellab{5} & \hyperlink{model.5}{software protection model} & \ref{sec:esp:risk_framing} \\
\modellab{3}  & \hyperlink{model.3}{asset value evolution model}    &      & 
\modellab{6} & \hyperlink{model.6}{actual threat model} & \ref{sec:esp:risk_framing}     \\

\end{tabular}
}

\caption{Models required in the proposed approach, with references to the discussions of their instantiation, if any.}
\label{tab:models}
\end{center}
\end{table}

\begin{table}[t]
\begin{center}

{\footnotesize \setlength\tabcolsep{3 pt}
\begin{tabular}{r l c |r l c}
Phase &  & & Phase & &  \\
\& No. & Method Name & ESP & \& No. & Method Name & ESP \\
\hline
1\ \ \ \methodlab{1} & \hyperlink{method.1}{primary asset description}           &   \ref{sec:esp:risk_framing}   &   
3  \methodlab{17} & \hyperlink{method.17}{mitigation deployment}                &   \ref{sec:esp:risk_mitigation:deployment}   \\
1\ \ \ \methodlab{2} &  \hyperlink{method.2}{software analysis tools}      &    \ref{sec:esp:risk_framing}  &   
3  \methodlab{18} & \hyperlink{method.18}{mitigation validation}                   &      \\         
1\ \ \ \methodlab{3} & \hyperlink{method.3}{secondary asset description}      &    \ref{sec:esp:risk_framing}  &   
3  \methodlab{19} & \hyperlink{method.19}{\softprot impact estimation}           &   \ref{sec:esp:risk_mitigation}   \\         
1\ \ \ \methodlab{4} & \hyperlink{method.4}{secondary asset identification algorithms}     & \ref{sec:esp:risk_framing}     &   
3  \methodlab{20} & \hyperlink{method.20}{single-pass mitigation decision making}                      &   \ref{sec:esp:risk_mitigation:hiding}   \\         
1\ \ \ \methodlab{5} &  \hyperlink{method.5}{requirement description}       &   \ref{sec:esp:risk_framing}   &   
3  \methodlab{21} & \hyperlink{method.21}{iterative mitigation decision making}     &      \\         
1\ \ \ \methodlab{6} &  \hyperlink{method.6}{export models of supported protections}   &   \ref{sec:esp:risk_framing}   &   
3  \methodlab{22} & \hyperlink{method.22}{asset hiding}     &   \ref{sec:esp:risk_mitigation:hiding}   \\         
2\ \ \ \methodlab{7} & \hyperlink{method.7}{threat analysis}      &    \ref{sec:esp:risk_assessment}  &   
3  \methodlab{23} & \hyperlink{method.23}{\softprot selection optimization}      &   \ref{sec:esp:risk_mitigation:optimization}   \\         
2\ \ \ \methodlab{8} &  \hyperlink{method.8}{threat impact estimation}    &   \ref{sec:esp:risk_assessment}   &   
3  \methodlab{24} & \hyperlink{method.24}{\softprot select search space pruning}       &   \ref{sec:esp:risk_mitigation:optimization}   \\         
2\ \ \ \methodlab{9} &  \hyperlink{method.9}{risk prioritization}    &    \ref{sec:esp:risk_assessment}  &   
3\  \methodlab{25} & \hyperlink{method.25}{cookbooks with \softprot recipes} &      \\         
2\  \methodlab{10} &  \hyperlink{method.10}{defender's analysis toolbox execution}   &  \ref{sec:esp:risk_assessment}    &   
3\  \methodlab{26} & \hyperlink{method.26}{driving the \softprot tool}   &   \ref{sec:esp:risk_mitigation:deployment}   \\         
2\  \methodlab{11} & \hyperlink{method.11}{incremental attack path enumeration}     &     &   
4\  \methodlab{27} & \hyperlink{method.27}{ risk analysis updating}     &      \\         
2\  \methodlab{12} & \hyperlink{method.12}{incremental threat analysis}    &      &   
4\  \methodlab{28} & \hyperlink{method.28}{application exposure monitoring}    &      \\         
2\  \methodlab{13} & \hyperlink{method.13}{transparent threat analysis reporting}    &   \ref{sec:esp:risk_assessment}   &   
4\  \methodlab{29} & \hyperlink{method.29}{monitoring risk framing input evolution}    &      \\         
2\  \methodlab{14} & \hyperlink{method.14}{risk monetisation}    &      &   
4\  \methodlab{30} & \hyperlink{method.30}{monitoring running applications}      &   \ref{sec:esp:risk_monitoring}   \\         
2\  \methodlab{15} & \hyperlink{method.15}{OWASP risk rating methodology}    &      &   
4\  \methodlab{31} & \hyperlink{method.31}{monitoring communication of running apps}                  &  \ref{sec:esp:risk_monitoring}    \\         
3\  \methodlab{16} & \hyperlink{method.16}{mitigation decision making}                  &   \ref{sec:esp:risk_mitigation}   &
4\  \methodlab{32} & \hyperlink{method.32}{user experience evaluation}             &      \\         
\end{tabular}
}

\end{center}
\caption{Methods in the proposed approach's phases, with references to the discussions of their instantiation, if any.}
\label{tab:methods}
\end{table}

We also highlight \introopenissue{open issues that are research challenges} and discuss where we think \introsota{existing state of the art can serve as a foundation}, in some cases by pointing out \introresearchdir{potentially useful research directions to find solutions}. We present \introrecommendation{recommendations and requirements}, in particular \introautomationreq{automation requirements}, and we highlight aspects on which different stakeholders need to perform \introfuturework{future standardization and engineering work} (as opposed to research). 


\subsection{Risk Framing}
\label{sec:framing}
In this phase of the approach, one defines the context in which a risk analysis will be performed. For the case at hand, one defines the relevant software targets, their assets and security requirements, potential attacks, available {\softprot}s, and \sdlc requirements. To enable standardization, a common vocabulary needs to be established that covers all possible constructs and models to describe all relevant scenarios. This needs to be unambiguous and formalized such that automated support tools can be engineered. 
\futurework{Provisioning the complete vocabulary to describe the risk frame} is, of course, out of reach here. That will instead need to be done in a larger document that results from a community effort. 
\sota{The meta-model of Basile et al.\ can serve as a starting point for modelling all the relevant constructs and their relations~\cite{reganoMeta}}.

\subsubsection{Assets} 
\label{sec:framing:assets}
A first task for a case at hand is to determine which assets are \emph{potentially} relevant. This is needed for all the potential assets known a priori, i.e., in the original application, in already deployed {\softprot}s, if any, or in any of the {\softprot}s that might later be deployed in the mitigation phase. 

The \construct{primary assets} are static and dynamic software elements of which a \mate attacker might violate security requirements because they have value for the attacker or the vendor: monetary value, public image, customer satisfaction, bragging rights, etc. Examples are secret keys or confidential data embedded in applications, algorithms that constitute valuable intellectual property or trade secrets, multiplayer game logic that needs to remain intact to prevent cheating (e.g., see-through walls, use aim-bots, or show full world maps), and authentication checks that need to remain in place. These assets are the primary targets of \mate attackers. They cover a range of abstraction levels and granularities corresponding to a range of code and data elements (functions, variables, global data, constants, etc.). For example, an algorithm can be large and expressed in abstract terms, while a secret encryption key to steal is merely a string of bits. Primary assets are already present in the vanilla, unprotected software.

The \construct{secondary assets} are software elements that attackers might target on their \construct{attack path} (i.e., the sequence of executed \construct{attack steps}) towards the primary assets. Attackers consider these elements as mileposts on their way to their primary targets. Secondary assets can be \construct{attack pivots} (a.k.a.\ hooks) in the vanilla software, but they can also be artifacts or fingerprints of injected {\softprot}s that attackers need to overcome. An example pivot is a ciphertext buffer containing high-entropy data, which an attacker might first try to identify with statistical dynamic analysis. Once the buffers have been identified, the attacker might pivot to the program slices that produce the buffers' data, and in those slices they can obtain the secret keys. An example of an injected {\softprot} is an integrity check. A gamer that wants to alter the speed with which he can move around in the virtual game world might first have to undo or bypass the integrity check. 

\recommendation{The distinction between primary and secondary assets should not be strict.}  For example, a cryptographic key that protects one movie might be a secondary asset if the attacker tries to steal one movie. A similar key that serves as a master key for all movie encryptions is clearly a primary asset. Moreover, \softprot vendors consider the {\softprot}s supported with their tools as primary assets that they do not want to be reverse-engineered easily. While those {\softprot}s protect the primary assets of their customers' software, they are the primary assets of the \softprot vendors. Should attackers learn how to attack or circumvent them automatically, their value goes down the drain.

The deployment of some {\softprot}s requires one to describe the relationship between assets and non-asset program elements. This is the case when {\softprot} transformations applied to the code of assets require other non-asset code to be transformed with it to conserve the program semantics. When deploying a \softprot on only the assets, this should not make those assets stand out to the attacker, e.g., because the entropy of encrypted data or obfuscated code is much higher than that of plain data or because the protection introduces recognizable fingerprints. To increase the attacker's effort needed to localize them, one can deploy the same {\softprot}s on non-asset code, as proposed by Regano et al.~\cite{reganoL2P}. Furthermore, to decide which {\softprot}s can be deployed conservatively, it might be necessary to analyse the whole application and model it. In short, an \model{application and asset model} is needed to describe the wide range of software elements that form the target application, including the elements of assets and non-assets, and the relevant relations between them. The \sota{application meta-model of Basile et al.} can provide a useful starting point~\cite{reganoMeta}, but it definitely needs to be refined, as it currently only captures coarse-grained relations such as call graphs. 

Multiple methods need to be considered for instantiating a concrete application model. Obviously, a \method{primary asset description method} is required to let a user identify and describe their primary assets, preferably at a high level of abstraction, such as with \sota{source code annotations}~\cite{D5.11}. Next, \automationreq{}\method{software analysis tools} need to map those descriptions onto the corresponding lower-level software elements (e.g., onto corresponding assembly operations) and extract the structure and relevant properties of the software. Such tools are already used in all \softprot tools we know of, both commercially and in research. If the \softprot decision support tools cannot identify secondary assets themselves, a \method{secondary asset description method} is required to let a user describe the secondary assets and how they relate to primary assets. Alternatively, we foresee that \automationreq{}\method{secondary asset identification algorithms} can be developed to automate their identification. Such algorithms would be executed in the later risk assessment phase, but in the framing phase, the necessary knowledge needs to be modelled in the form of \model{secondary asset models} that describe what technical attributes of software elements allow attackers to exploit them as mileposts. An example is the already mentioned buffers that contain high-entropy data. Precisely the fact that some buffer holds such data makes it a potential milepost. \openissue{The design of such secondary asset models is an open issue.} Note that those models would not need to be recreated from scratch for every application. Instead, they would be reusable and grow over time as new types of secondary assets are considered.

As {\softprot} aims to delay attacks rather than prevent them, we need \model{asset value evolution models} to describe the evolution of their value over time, including the \construct{attack time frame} in which assets have value as well as the impact a successful attack can have on a business model. This includes the \construct{renewability of assets}, i.e., how easy it is to replace software and assets to reduce the impact of successful attacks. For modelling this evolving relationship between business value and assets, we expect that companies can use \sota{their existing asset valu\-ation models.}


%
\futurework{To enable asset risk framing in a standardized manner, stakeholders first need to join forces to draft a taxonomy of possible assets and their features.} 
A starting point can be \sota{Wyseur's list of assets} in the form of private data, public data, unique data, global data, traceable code/data, code, and application execution~\cite{D1.02}. Another starting point can be Ceccato et al.'s \sota{taxonomy of code and data elements} that MATE attackers considered in their experiments~\cite{emse2019}.

\subsubsection{Security Requirements} 
The \construct{primary security requirements} of assets are often the \construct{non-functional requirements} of confidentiality and integrity. These come in different forms, levels of abstraction, and granularity. Their scope differs from that in other domains, so their classifications can not be trivially reused. For example, \mate integrity requirements can include constraints on where or how code is executed, that at any point in time at most one copy of a program is running, and that certain program fragments are not lifted and executed ex-situ. In addition, there might be non-repudiation requirements. For example, unauthorized copies must be detected upon execution.  

\recommendation{For different phases in the software \sdlc, different requirements may hold, and different types of attack activities may need to be mitigated}, such as in the \construct{attack identification phase} versus the \construct{attack exploitation phase}. Some requirements may be absolute, such as a master key that should never leak; others may be time-limited, such as a key to a live  event that should remain secret for 5 minutes; still, others may be relative and economical, such as that running many copies in parallel undetected should cost more than licensing them.

Assessing whether non-functional requirements can be guaranteed is hard in practice because of the \mate attackers' white-box access. \construct{Secondary security requirements} can help frame possible risks. These can be (i) non-functional requirements for secondary assets; (ii) \construct{functional requirements} that are easier to check but of which the mere presence in itself provides few guarantees, such as the presence of a copy-protection mechanism; (iii) \construct{assurance security requirements} that minimize the risk that relevant aspects are overlooked; and (iv) what we will call \researchdir{}\construct{protection policy requirements}. The latter relates to worst-case assumptions about attacker capabilities, such as assuming that the mere presence of some features suffices to enable certain attacks. Such assumptions can compensate for the lack of proper evaluation of primary requirements. For example, a lack of stealth resulting from easily identifiable invariants in injected {\softprot}s hints for potential \construct{weaknesses} vis-\`a-vis certain attacks~\cite{yadegari}. Protection policy requirements then require that elements with certain features are not present at all or meet certain requirements, such as statistical properties. This is similar to security policies in the domain of remote exploitation, where, e.g., code pointer integrity is a policy about handling code pointers that can ensure that indirect control flow cannot be hijacked by exploits~\cite{CPI}.

In the risk framing phase, the task for a case at hand is to determine and describe the security requirements for all assets and \emph{potential} weaknesses identified as relevant. A \method{requirement description method} is needed for the user to describe their primary and part of their secondary requirements, using a requirement taxonomy. One option is \sota{to annotate the source code}~\cite{D5.11,D5.13}. \futurework{Standardizing a taxonomy requires a community effort.} \openissue{How to model protection policy requirements} is an open issue. It is closely related to the secondary asset model discussed in the previous section; the necessary models will hence best be co-designed.

\subsubsection{Attack Models} 
\mate risk management needs to consider a range of potential attacks described in an \model{attack model}. 
This needs to cover attackers with different levels of \construct{attack resources} and \construct{attack capabilities}: money, expertise, available tools, etc. The latter involves a range of methods and evolves over time, so a \futurework{living catalog} is needed. \openissue{We currently do not know what level of detail will produce the best results}, so both more generic attack methods and tool usage scenarios (e.g., disassembling code) and very concrete ones (e.g., using the IDA Pro 8.0 disassembler) need to be supported. As the goal of \softprot is to delay attacks, \recommendation{not only the feasibility of successful attacks is to be covered, but also the potential effort involved}, possibly including what attackers would probabilistically waste in unsuccessful attack strategies. 

While research has shown that attackers commonly waste time on unsuccessful attack steps in real attacks~\cite{emse2019}, \openissue{it is unclear whether useful attack models can build on} \construct{worst-case scenario assumptions}. Examples are attackers being served by an oracle always to choose the right attack path, and analysis tools producing results with ground-truth precision. For example, locating the code of interest is an important, time-consuming attack step that cannot simply be assumed to be performed effortlessly using an oracle~\cite{ReMind}. Doing so would imply that increasing the stealth of {\softprot}s is not useful, which experts certainly reject. 

For each potential attack step, the attack model needs to encode which features of software elements are \construct{attack-enabling features}, \construct{attack-preventing features}, and \construct{attack effort determination features}, i.e., that enable or prevent an attack step, or that significantly affect the required time and effort of an attack step, as well as the \construct{attack likelihood of success features}. An example is the presence of certain secondary assets. These features might include features of the software under attack, the environment in which attacks can be performed, but also knowledge obtained by the attacker. \openissue{The best abstraction levels to consider are an open question.} 

The same holds for the \construct{software protections} and a set of (quantitative) \construct{protection strength metrics} that can be used in later phases to estimate the effort/time/resources that attackers will need to invest in the attack steps in scope. Depending on the maturity of a decision support tool, that set may have to be selected manually during the risk framing. As discussed in Section~\ref{sec:challenges}, there currently is no widely accepted set of metrics. Many proposals~\cite{Anckaert2007,D4.06} have been made on features that should be measured (e.g., control flow complexity) and on concrete metrics for doing those measurements (e.g., cyclomatic complexity~\cite{mccabe} or code comprehension~\cite{Tamada2012}). \sota{Those proposals on metrics can serve as starting points}, but \openissue{more empirical research is needed on top of existing work~\cite{ReMind} to determine which metrics are valid under which circumstances and for which purposes.} \sota{RevEngE} by Taylor and Collberg seems to be a good approach for enabling more productive research of human attack activities~\cite{RevEngE}. In the context of the Grand Reverse Engineering challenge\footnote{\url{https://grand-re-challenge.org/}}, their data collection software is not only used to analyze attacks on randomly generated programs but also on purposely designed MATE challenges, which allows studying the relations between human attack effort and metrics. \sota{For automated attack tools, such as symbolic execution or black-box deobfuscation, the framework proposed by Banescu et al.\ can be a  starting point~\cite{Ban17}.} Because that framework relies on \ml, thus requiring evaluations on many samples, it is not suited for manual attack activities.  

In the risk framing phase, the task for a case at hand is to determine the attack model, i.e., the combinations of the mentioned attributes that potential attackers in scope might \emph{potentially} have. 
Existing models from network security risk analysis cannot be reused. \recommendation{\mate attack modelling needs to include manual tasks and human comprehension of code}, which are not considered in network security. 
For example, in network security, the development of zero-day exploits (using tools also found in the \mate toolbox) is handled as an unpredictable event, which side-steps the complexity of analysing and predicting human activities. \sota{This entirely prevents the use of existing assessment models developed for the network security scenario}. 

Some \sota{studies document how \mate attackers operate in practice}~\cite{ceccatoTaxonomy,emse2019,ReMind,D1.02attack}. Together with \sota{numerous blogs and case studies by reverse engineers}, such as those by Rolles~\cite{virtualizationAttack}, they can help to determine an appropriate attack model. \sota{Existing MATE attack taxonomies} can also be built upon to enable users to formulate attack models for their cases~\cite{MATEtaxonomy,ahmadvand2018,banescu2017tutorial}.

\subsubsection{Software Protections}
A \model{software protection model} is needed to describe in a unified manner the wide range of {\softprot}s that a user's tools might support. This model needs to include at least possible limitations on \construct{applicability} and \construct{composability}, be it for \construct{layered SP deployment} to protect each other or to exploit \construct{synergies} between multiple {\softprot}s; the security requirements that they help to enforce; (measurable) features or limitations they have that can enable, slow-down, ease, block, or otherwise impact potential attacks, on the {\softprot}s themselves but also on the assets they are supposed to protect; how big those impacts are on the potential attacks; and potential implementation weaknesses including how they can fail to meet protection policy requirements and become (easily) attackable assets themselves; etc. The link to validated (but as of yet still missing) metrics mentioned above is clear, and the impact that deployed SPs have on metrics used to asses attack effort, i.e., the \construct{potency}, \construct{resilience}, and when relevant the \construct{stealth} of potentially deployed SPs obviously also needs to be modelled. 

The \softprot model needs to capture the costs of using a \softprot. This can include the direct monetary costs of {\softprot} tool licenses, but also indirect costs such as having to budget for more security servers or having a longer time to market, or any other  cost that might follow from changes to the \sdlc.    

The potential overhead of all available {\softprot}s needs to be known w.r.t.\ run time, latency, throughput, size, ... This is critical because many applications have a little overhead budget when it comes to responsiveness, computation times, etc. In part, the performance impact depends solely on a {\softprot} itself, such as the (constant) time or memory required to initialize it. The impact can also depend on how a {\softprot} is deployed. For example, whenever a \softprot requires the injection of a few instructions into code fragments, the resulting overhead will depend heavily on how frequently executed those fragments are. \recommendation{Multiple ways for expressing the potential cost of {\softprot}s are hence needed.}

In the risk framing phase, the user needs to determine which combinations of {\softprot}s can \emph{potentially} be deployed to mitigate risks, given the available {\softprot} tools. For automating the later phase of risk mitigation, \automationreq{the used \softprot tool should be able} \method{to export a model of all discussed features of all SPs it supports}, such that a decision support tool can import that model and such that the tool user does not have to provide the information manually. Therefore, the SP tool vendor is responsible for instantiating the SP model of their tool. \futurework{This obviously requires tool vendors and other SP stakeholders to agree on a standardized taxonomy of SPs and their relevant features.}  To model the available composability, the \sota{finite state automata} proposed by Heffner and Collberg to model pre/post-requirements, pre/post prohibitions, and pre/post suggestions for combinations of {\softprot}s are an interesting idea~\cite{heffner2004obfuscation}. 

\subsubsection{Software Development Life Cycle Requirements}
\label{sec:framing_sdlc}
{\softprot}s come with side-effects, such as slowing down software, making it bigger, making debugging harder, requiring changes to distribution models, requiring certain scalability on the side of secure servers, etc. Taking the time to decide on {\softprot}s, possibly iteratively with the involvement of experts and time-consuming human analysis, also affects the time to market.

Hard and soft \construct{constraints} need to be collected in terms of quantifiable overheads/costs in all possible relevant forms, and with respect to compatibility with \construct{\sdlc requirements}. Different constraints might apply to different parts of a program. For example, in an online game or a movie player, the launching of the game or player might have a large overhead budget, while during the game or movie real-time behavior is critical. 

For all available {\softprot}s, later phases of the risk analysis will need to estimate the impact on the relevant costs and \sdlc. It is, therefore, necessary to obtain all relevant \construct{profile information} on the software, including execution frequencies of all relevant code fragments. 

An important complication occurs when the vendors of \softprot tools (hereafter named \softprot vendors) and users of such tools (hereafter called application vendors) do not trust each other. Both parties often put severe constraints on how the \softprot tools are deployed and on the amount of information they exchange. A \softprot vendor will typically not be very forthcoming about the weaknesses or internal artifacts of the supported {\softprot}s and disallow reverse engineering of them, while the application vendors do not want to share too many details or code with the \softprot vendor. Consequentially, only illegitimate attackers will get white-box access to the protected applications in which {\softprot}s and original assets are interwoven as discussed in Section~\ref{sec:challenges}. If the experts performing the risk management lack white-box access to all available {\softprot}s and to the protected application, this will have a tremendous impact on the methods and data that can be used during the risk assessment and risk mitigation phases that target attackers with white-box access. \recommendation{This lack of white-box access by the defenders obviously needs to be documented, and the potential impact thereof needs to be assessed} during the risk framing.


In addition, aspects of the \sdlc relevant to the monitoring phase (that will be discussed later) need to be framed, such as \construct{connectivity} and \construct{updatability}. Whether an application will always be online, occasionally connected, or mostly offline impacts which online {\softprot}s and which monitoring techniques can be deployed. So does the ability to let application servers such as video streaming servers or online game servers interact with online security services such as a remote attestation server. Likewise, it is important \recommendation{to document whether updates can be forced upon users and to what extent the vendors can synchronize users' updates}.

Finally, \construct{limitations to the environment} in which software will be distributed and executed need to be documented. For example, Android supports fewer OS interfaces for debugging, and some device vendors limit what applications can do after installation, such as iOS's limitation on downloading binary code blobs post-installment. Such limitations clearly affect the types of {\softprot}s that can be deployed, so they need to be included in the risk framing.

\automationreq{To avoid the need for costly human expertise and manual intervention in the next process phase, as much as possible information discussed above needs to be formali\-zed, such that tools can reason about them in the subsequent phases.} As already noted at the beginning of Section~\ref{sec:framing}, this obviously requires a standardization effort by the community to create a standard vocabulary and taxonomies that cover all constructs and models to be documented in the risk framing phase.



\subsection{Risk Assessment}
\label{sec:assessment}

In the discussion of risk framing, the term ``potential" occurs frequently, because in that phase all forms of knowledge are still considered in isolation, including potential \softprot weaknesses, application features, \softprot tool capabilities, and attacker capabilities. In the risk assessment phase, one assesses how they interact for the case at hand by determining which of all potential risks actually manifest themselves in the software at hand. First, a \method{threat analysis} needs to identify the \construct{actual threats} starting from an analysis of the assets and their intrinsic weaknesses, as well as from attack strategies and their technical attributes that impact their feasibility.
Then a qualitative, semi-qualitative, or preferably quantitative \method{threat impact estimation} needs to be performed to identify the \construct{actual risks}, and a \method{risk prioritization} needs to be done.

\subsubsection{Identification of the Actual Threats} 
\label{sec:identification_threats}

This phase aims to determine a list of attacks that could succeed on one or more of the application's assets by violating their security requirements.
This phase therefore consists of a detailed threat analysis that outputs a \model{actual threats model} that describes those analyzed attacks deemed feasible within the assets' relevant attack time frames, i.e., the actual \construct{attack surface} and the \construct{attack vectors} on it (e.g., exploited pivots and weaknesses), the \construct{attack paths of least resistance} among them, the levels and amounts of expertise, effort, and resources attackers need to mount those attacks, the damage caused by exploitation, etc. 
For each attack path contributing to the major threats, \recommendation{the weaknesses and secondary assets used by attackers as pivots need to be included, as well as the used assumptions, such as worst-case-scenario considerations or parameters that are unknown in practice}. Reporting this information in an actual threat report is necessary to enable confidence in the outcome of the assessment. 

Critically, \recommendation{the enumeration and assessment of feasible attack steps must be performed on both the attack identification phase and the attack exploitation phase}. The former takes place in the attacker's lab on their infrastructure, the latter more often takes place on other users' devices. 

Several open issues need to be addressed to perform this task correctly. First, \futurework{standardization should produce a more precise approach and methodology for defining the \mate threat model, the attack surface, and attack vectors.} The latter includes the information attackers can extract from the target software. Assets can be attacked with different strategies, in which attackers rely on automated tools and analyses to collect and exploit information about the software and to represent the software in structured representations. A range of \construct{analysis tools} and techniques are applicable, all with their own strengths and limitations, including static, dynamic, symbolic, and concolic analyses. 
Knowing the attacker's goals and tools is the starting point for identifying and enumerating the possible attack paths. This knowledge includes the kinds of analysis results that the different tools can produce, i.e., \construct{software features} such as taint information, profiles, data, and control flow dependencies. It also includes the software features those analyses depend on to produce their results, their weaknesses, limitations, and precision. 

In this phase \method{the defender hence needs to deploy their own analysis toolbox} to determine the features of the  primary assets and related application elements that can have an impact on the feasibility of attacks because they enable, prevent, slow down, or otherwise impact attacks. This includes \recommendation{checking whether the protection policy requirements formulated in the risk framing phase are violated}. It also \recommendation{needs to be done for all potential weaknesses} that were identified in the framing phase, such as invariants or fingerprints in the code that might facilitate certain attack vectors. Moreover, \recommendation{the set of actually present secondary assets needs to be determined} to identify the presence of features that make them pivots for attackers towards the primary assets. \automationreq{Obviously, most if not all of the analyses in the toolbox should be applied automatically.}

While we are convinced that such defender toolboxes can produce most of the necessary information for enumerating feasible attacks, a number of research questions are open. 
For example, \openissue{how can the formal pieces of information extracted by the tools be used to precisely identify the viable attack paths?} In particular, when attackers need to resort to manual efforts, that is not easy to formalize. 
\openissue{How do we then assess the required effort and likelihood of success?} 
\openissue{To what extent can automated analysis with a defender toolbox suffice to avoid the need for actual penetration testing involving human experts?} 


It is also an open question \openissue{how fine-grained or concrete the enumeration of con\-sidered attacks paths and their attack steps needs to be} and how their attributes are to be aggregated.
Since the assessment must drive the mitigation, the generated information must be rich enough for the mitigation decision makers. 
Therefore, to some extent, the answer to the above question will depend on the goal of the assessment. This can be a semi-automated or fully automated mitigation phase. In the latter case, assessment information must be extensive and accurate, as an automated decision support system cannot rely on human intuition and experts' past experience.


The identification of attacks with an analysis toolbox requires \recommendation{white-box access to the application code}. In case this is not possible, e.g, because of \sdlc requirements discussed in Section~\ref{sec:framing_sdlc}, \recommendation{alternative sources of information about the different integrated components need to be considered}, such as \construct{third-party-provided incomplete analysis} reports, i.e., partial analysis reports provided by the involved parties. Alternatively, and as long as the discussed enumeration approach cannot completely replace human expertise, the inclusion of results of penetration tests performed by  red teams could be considered.  
In short, \openissue{the threat analysis needs to be able to take into consideration a wide range of information sources and forms.}

For the scalability and practical use of a software threat analysis process, another open issue is \openissue{} \method{incremental attack path enumeration}, i.e., how to update and maintain the attack path enumeration without repeating a full analysis from scratch when any of the involved aspects evolve while the application is still being developed, be it the application itself, the \softprot tool flow, the attackers' tool boxes, etc. Especially if the attack enumeration involves human expertise, a solution in the form of \automationreq{}\method{incremental analysis} is critical. 

The current state of the art still requires such human expert involvement. Past research aimed to \sota{auto\-mate the attack discovery with} abductive logic and Prolog~\cite{basileOTP,reganoProlog}. That suffers from computational issues, since generating attack paths as sequences of attack steps causes a combinatorial explosion and requires massive pruning.
With the pruning by Regano et al.~\cite{reganoProlog} only high-level attack strategies can be generated, which often do not contain enough information to make fine-tuned selections among similar {\softprot}s. For example, they allow determining the need for using obfuscation but do not provide hints for selecting among different types of obfuscation.

\researchdir{\ml} might be useful to synthesize attack paths from attack steps more effectively~\cite{10.1145/2229156.2229157}.
Moreover, \researchdir{}\sota{methods for exploit generation}~\cite{brumley_apeg,angr} that automatically construct remote exploits for vulnerable applications could be investigated to determine \mate attack paths automatically. 
They will certainly need modifications, as finding exploitable vulnerabilities is rather different from finding \mate attack paths.
For example, in the \mate threat analysis, for each identified attack path \recommendation{defenders need to estimate the likelihood of succeeding as a function of the invested effort, attacker expertise, time, money, and luck in trying the right strategy first or not}, etc. All of that is absent in the mentioned automated exploit generation.  

Regarding automation, we think the identification and description of primary assets cannot be automated, as those depend on the business model around the software. They can hence not be determined by only analysing the software. By contrast, \automationreq{the identification of secondary assets, as mentioned in Section~\ref{sec:framing:assets}, as well as the discovery of attack paths and the assessment of their likelihood, complexity, and other risk factors, should be prime targets for automation}.

Even if full automation is out of reach because parts cannot be automated or do not produce satisfactory results, automating large parts of the threat identification phase will already have benefits. It will reduce human effort, thus making proper risk assessment cheaper and hence more accessible, and it can raise awareness about identified attack strategies, thus making the assessment more effective. \researchdir{A gradual evolution from a mostly manual process, over a semi-automated one, to potentially a fully automated one, is hence a valuable R\&D goal}. We stress that in order to succeed, automated tools should then not only provide the necessary inputs for later (automated) phases of the risk management, \automationreq{they should also enable experts to validate the produced results to grow confident in the tools}, by \method{providing a transparent report on the performed threat analysis.}  Section~\ref{sec:workflow} will present a tool that, although rather basic, achieves just that.
\sota{For presenting the threats to human experts, different formats have been proposed in the literature, including attack graphs~\cite{attack_graphs} and Petri Nets~\cite{petri_nets_attacks}}.

\subsubsection{Evaluating and Prioritizing Risks}
\label{sec:req:prioritization}

The \recommendation{risk assessment report must indicate the consequences that exploitation of an identified actual threat may have. It must produce an easily intelligible value or score associated with all the risks to all assets}. 
Since the objective of the report is prioritizing the risks to drive the mitigation phase, \recommendation{it must not only consider the direct value of the violated primary assets, but also the side effects}, like impact on the business reputation or market share losses.

Furthermore, \recommendation{it may consider the likelihood that attack\-ers are interested in executing the identified threats} because of different expected {\roi}s. For example, an attack path that offers a lot of potential gains for the attacker might be less attractive when it comes with a high probability of being detected and having to face legal consequences.

When outcomes from the impact analysis are available in proper form, our feeling is that this phase has no peculiarities compared to risk analysis in other fields. 
Models and methods can therefore be adopted from existing literature to build a system that allows the consistent evaluation of the impact. 
As a promising option, we consider \researchdir{}\sota{}\method{risk monetisation}~\cite{doerry2015monetizing}, the process of estimating the economic loss related to risk and the \roi of mitigation activity. This eases reporting to higher management and is general enough to work for every asset type, including software assets.
\researchdir{}\sota{Investigating the aspects of the }\method{OWASP risk rating methodology} could also yield interesting results  that might work in the \mate context~\cite{OWASPrisk}. 
Automation support for the available options can then obviously also be reused, possibly after some adaptations.

\subsection{Risk Mitigation}
This phase comprises two parts: first \method{mitigation decision making}, and next there is \method{implementing} and \method{validating} the decisions.  

\subsubsection{Mitigation Decision Making}
\label{sec:decision_making}
\recommendation{Risk mitigation requires the defenders to evaluate how the deployment of combinations and configurations of {\softprot}s will affect the high(est) risk attack paths}. 

Ideally, this evaluation can be done through \method{\softprot impact estimation} without having to actually deploy the considered {\softprot}s and having to measure their effect. This is a major difference from the risk assessment phase, which relied heavily on measurements. \openissue{How precise the estimations need to be to enable a sufficiently precise comparison of} \construct{residual risks} is an open question. We consider two possible approaches. 

First, we consider \method{single-pass mitigation}. \researchdir{This builds on an assumption that estima\-tions are accurate enough to determine the best possible combination of {\softprot}s without additional measure\-ment.} A human or tool then first determines the \construct{most protective selection}, i.e., the combination and configuration of SPs that achieves the minimal residual risk while not violating hard constraints. Next, one selects \construct{alternative protection targets} that trade off some of the residual risks for other aspects, such as lower performance penalty. For each alternative target, one then again selects the best target-specific {\softprot}s and estimates the delta in residual risk and in other relevant aspects over the selection that yielded the minimal residual risk. Finally, one then chooses between the most protective selection and the alternatives. This human decision will typically involve {\softprot} experts, application architects, and managers familiar with the business strategy. Given the complexity of {\softprot} as discussed before, we consider such a decision making process not automatable at this point in time, nor in the near future. 

The alternative is \method{iterative mitigation}. \researchdir{This approach, which is familiar to practitioners in the industry, adds additional {\softprot}s iteratively in a layered fashion.} The assessment and mitigation phases are not executed once, but alternated over multiple rounds. In each \construct{mitigation round}, an assessment is followed by mitigation. In the first round, the risk assessment is done on the vanilla application. In later rounds, the assessment is performed on the application protected with all {\softprot}s selected in previous rounds. During such later assessments, measurements are performed on already selected and deployed {\softprot}s. This works around the lack of precise enough estimation methods as needed for the first approach. 
It also eases the handling of novel risks introduced by deployed {\softprot}s, such as when the location of non-stealthy {\softprot}s might leak the location of assets.

In each round, the mitigation adds an SP layer consisting of a few additional {\softprot}s to the ones already selected in previous rounds. In each round, different combinations of {\softprot}s can be proposed that offer different risk reduction and cost trade-offs. Humans will then again select one combination and continue to the next round, or stop once the whole cost budget is consumed or no more significant risk reduction is achieved. In each round, different constraints can be imposed that limit the {\softprot}s considered in that round, and the set of {\softprot} is chosen that offers the best potential to reduce the residual risk. Estimating the reduction potential rather than the immediate reduction in each round allows for taking into account a priori knowledge about the fact that some {\softprot}s have the potential to become much stronger after additional rounds corresponding to additional layers are deployed, while other {\softprot}s cannot become stronger because of a lack of synergies.

An example of constraints evolving between rounds is that in the first rounds {\softprot}s might only be deployed on assets, while in later \method{asset hiding} rounds, non-stealthy {\softprot}s can be added for non-assets to avoid that protected assets stand out because of \softprot fingerprints. Our \poc contains such an asset hiding step, albeit in the same round as the asset protection step, i.e., without performing measurements in between, as will be detailed in Section~\ref{sec:esp:workflow:hiding}.

The iterative approach is more realistic for several reasons. The humans making decisions in each round can make up for deficiencies in the existing tool support and formalized knowledge, and they can build more confidence in the outcomes of the mitigation process. Secondly, measurements are performed in each round, which again allows for more confidence in the outcomes. 

Automation poses the most severe constraints on the mitigation task.  
\method{Optimizing the selection of of {\softprot}s} must comply with computational constraints. \automationreq{In most usage scenarios, optimization models must return results within minutes or hours.} 
Given the large search space to explore, this requires ad hoc \method{search space pruning} methods that prune less relevant combinations efficiently. In some usage scenarios, optimization models returning far-from-optimal results quickly are acceptable, such that the time-to-market requirements of a software launch can be met while spending more time to find better {\softprot} combinations for later updates.\footnote{Anonymously, \softprot suppliers confirm to us that for many of their customers the norm is weak implementation at first because security/protection is not on the feature list from product management, and then complaining when things get broken, after which the supplier needs to help out. Obviously, they prohibit us to document concrete cases.}

\recommendation{Within one round of decision making, the optimization process should be driven by at least the potency of the selec\-ted \softprot combination and by estimating the protected software's performance. Ideally, resilience is also considered.} Current methods for estimating the potency, resilience, and (to some extent) overheads are not usable for automatic decision support, as they require the deployment of the {\softprot}s to perform a measurement. Given the time and resources needed to apply {\softprot}s on non-toy programs to measure objective metrics and run-time overheads, an optimization process that requires measurements instead of estimations would only consider a very limited solution space, which would make the optimization process useless.

\openissue{Estimating the strength (and overhead) of layered {\softprot}s} is really hard, as their code is highly interwoven.
Our work on \sota{estimating the potency of obfuscations~\cite{reganoMetric}} and on \sota{a game-theoretic approach to optimize the selection of {\softprot}s~\cite{ReganoPhd}} will be discussed in Section~\ref{sec:esp_optimization_approach}.
\researchdir{\ml can likely help to solve this difficult problem}, as already demonstrated for specific aspects of strength such as resilience against symbolic execution~\cite{Ban17}, but clearly need further research.

\openissue{Another open issue is that {\softprot}s have varying effects on attack success probability}, in particular when the security requirements are time-limited or relative. In some cases, the effects can be quantified in absolute terms, such as increased brute-force effort required to leak an encryption key from well-studied white-box crypto protection. In other cases, such as the delay in human comprehension of code that has undergone design obfuscations~\cite{collbergbook}, the effect is harder to quantify. When software contains different assets with different forms of security requirements, the relative value of different {\softprot}s hence becomes  difficult to determine, and hence the overall risk mitigation optimization becomes increasingly difficult.

\sota{There is a limited body of existing work available on the described decision support}, and it does not cover all necessary constructs, models, and methods. In industrial practice, companies provide so-called \method{cookbooks with {\softprot} recipes}. For each asset, users of their tools are advised to manually select and deploy the prescribed SPs in an iterative, layered fashion as long as the overhead budget allows for additional {\softprot}s. Automated approaches are either overly simplistic or limited to specific types of {\softprot}s, and hence only support specific security requirements. 
Collberg et al.~\cite{collberg1997taxonomy}, and Heffner and Collberg~\cite{heffner2004obfuscation} studied how to decide which obfuscations to deploy in which order and on which fragments given an overhead budget. So did Liu et al.~\cite{obf_optvialangmods,7985664}. They differ in their decision logic and in the metrics they use to measure {\softprot} effectiveness. Importantly, however, their used metrics are fixed and limited to specific program complexity and program obscurity metrics, without adapting them to the identified attack paths.
Coppens et al.\ proposed an iterative software diversification approach to counter a concrete form of attack, namely diffing attacks on security patches~\cite{coppens2013feedback}. Their work measured the performance of concrete attack tools to steer diversification and reduce residual risks. All of the mentioned works are limited to obfuscations. In all works, measurements are performed after each round of transformations, much like in the second approach we discussed above. 

To improve the user-friendliness of manually deployed \softprot tools, Brunet et al.\ proposed composable compiler passes and reporting of deployed transformations~\cite{10.1145/3338503.3357722}. Holder et al.\ evaluated which combinations and orderings of obfuscating transformations  yield the most effective overall obfuscation~\cite{obf_evaloptphaseord}. However, they did not discuss the automation of the selection and ordering according to a concrete program and security requirements.

\subsubsection{Actual Deployment}
\recommendation{In each mitigation round, the chosen {\softprot} combination needs to be deployed}, so {\softprot} tools need to be configured and run to inject the {\softprot}s selected so far. Ideally, this is completely automated. \automationreq{This requires tool interfaces that allow} \method{the decision support tools to drive the {\softprot} tools}. Providing such interfaces and enabling this automation would have significant benefits. Besides saving effort on manual user interventions, it would also skip the learning curve of configuring the used tool flows properly. Moreover, having such an integrated framework could pave the road for an open standard for an API for {\softprot}.

Following the deployment, \recommendation{it is critical to validate that the {\softprot} tools actually delivered as expected.} Were the selected {\softprot}s injected in the intended way? Do the injected {\softprot}s have weaknesses that were not expected? \openissue{How to obtain the necessary validation is an open question in some usage scenarios,} in particular when the deployment of the mitigation is executed by multiple parties that do not want to share sensitive information and do not provide white-box access to their software components.


\subsection{Risk Monitoring}

According to the NIST~\cite{nistSP800-37} risk monitoring includes ``assessing control effectiveness, documenting changes to the system or its environment of operation, conducting risk assessments and impact analysis, and reporting the security and privacy posture of the system.''
For \softprot, \recommendation{monitoring involves the continuous tasks to be performed once the software has been released} to track how the risk exposition evolves over time. 
This consists of two related activities: \method{updating the risk analysis}, and \method{monitoring the risk exposure of the released application}.

\subsubsection{Keeping the Risk Analysis Up-to-date}

Keeping the analysis up-to-date requires \method{monitoring how the inputs used in the three earlier risk analysis phases evolve over time and how that evolution affects the decisions made in those phases.} We can abstract these into monitoring the evolution of three different pillars of information: the information related to the assessments (\eg, new attacks, attack techniques, tool updates), the information related to {\softprot}s (\eg, updates, vulnerabilities, breaches), and the information related to the protected application. 
Of course, monitoring can then lead to the decision that a differently-protected version of the application should be released whenever any tracked changes lead to a re-evaluation of earlier decisions.

The \recommendation{monitoring of information related to {\softprot}s con\-cerns both attacks against existing {\softprot}s and newly developed {\softprot}s}. For example, when a complex attack technique (e.g., generic deobfuscation~\cite{yadegari}) is first presented in the academic literature, it might not be considered relevant during an original risk assessment because the attack is hard to replicate and its effectiveness has not been demonstrated on more complex pieces of software. However, when attackers later release a toolbox that automates the replication and publish a blog discussing how they used it to attack a complex application successfully, this should lead to a re-evaluation. Similarly, \recommendation{when new {\softprot}s become available with higher effectiveness against old or new attacks, or with lower overhead, this may lead to a re-evaluation}.


Similarly, \recommendation{monitoring must consider that information related to the application can itself evolve}. One example is where a company might decide that there are, in fact, additional assets in the program that need to be protected. This can happen both as a late realisation after deployment, but also in the case where the application itself evolves over time, by virtue of new versions being released with changes in functionality or structure. Another example is that the priorities in the company's value estimation can change over time. This would mean that the associated formulas for the risk analysis produce different values.

\subsubsection{Risk Monitoring of the Released Application}
\label{sec:monitoring_relaesed}

Next, one needs to \method{monitor how copies of the released soft\-ware are running on their users' premises}.
This can be achie\-ved by \method{monitoring the information that the protected application communicates to the vendors}. 
Such information may originate from a \sota{monitoring\-by-design \softprot such as reactive remote attestation~\cite{viticchie2016reactive}}, but also from communication with other online components that were not originally designed for online monitoring.
This is particularly the case when anomaly detection can link irregular communication patterns to unauthorized activities, such as running multiple copies in parallel or executing program fragments in a debugger in execution orders or frequencies not consistent with authorized uses. 
Such patterns can occur from communications present in the original applications, or from online {\softprot}s such as code \sota{renewability~\cite{renewability}} and \sota{client-server code splitting~\cite{barrierslicing}}. 
Importantly, the use of non-monitoring communication does not require the implementation of reaction mechanisms in the protected application to be effective. 

In many cases, \recommendation{it is advisable to analyse the data obtained with the monitoring}. The insights extracted can be helpful to respond to detected anomalies, for example, by letting the application server take action in case of discovered attacks, as well as to keep the risk analysis process up-to-date, for example, to re-valuate the threats and their likelihood. 

Finally, the vendor of the released application needs \method{user experience evaluation methods} to monitor whether the impacts of the deployed {\softprot}s  on the user experience and cost are in line with expectations or promises by the \softprot vendor. For example, if users start reporting usability issues or if online {\softprot}s lead to scalability issues, \eg because more copies are sold than originally anticipated, those evolutions might also warrant a revision of the risk mitigation strategy.

\if false
\section{Automation Requirements}

\changed{
It is on the mitigation task that automation poses the most severe constraints. 
First, optimizing the selection of the combination of {\softprot}s to apply in order to mitigate the risks must comply with strict computational constraints. In turn, this implies the solution of ad hoc optimization models that, given the very large size they may reach, must guarantee that results are obtained in useful time (minutes, hours, rarely days) and that pruning actually discards the less important combinations.

Moreover, an optimization process should decide based at least on the potency of the selected combination of {\softprot}s and on the estimation of the performance of the protected app (\eg user experience). Current methods for estimating the potency and overheads are not usable for automatic decision support, as they require the actual application of the {\softprot}s.
Given the time and resources needed to apply {\softprot}s on non-toy programs, measure objective metrics, and compute or estimate the overheads by running the protected applications, the optimization would only consider a very limited solution space, making an optimization process useless.

Therefore, methods to predict the potency and the overheads are needed when a combination of {\softprot}s is applied.
As a solution, we have applied ML techniques to estimate the changes in the objective metrics after the application of {\softprot}s. From the predicted metrics we compute the new potency. Also, our simplified overhead estimation model is based on predicted metrics and data about the {\softprot}s, nonetheless, a new model that uses ML is under development.

As the last part that would complement automatic decision support, we mention the automatic application of the selected {\softprot}s. This step would avoid the complexity of learning how to properly configure the application of {\softprot}s. Moreover, having such a framework would pave the road for an open standard for an API for {\softprot}.
}

{\color{brown}

\begin{itemize}

\item automatically identify the assets in the software is definitely outside the reach

\item 
identify the secondary assets to protect from an explicit specification of the primary assets in the software, \eg, by  using a sort of attacker model catalogue or as the output of automatic attack discovery. 
Secondary target: show defenders that they protect much more than just assets, focus attention on a larger important part of the code, not just the assets

\item automatically determine which attack/attack steps can be automated and when the manual effort is needed

\item ??? correctly model all the artifacts that tools produce (or may produce) to help attackers in mounting attacks. Associate some usability scores. 

\item automatically identify the attacks/attack paths, the resulting findings must be rich enough to be used for fine-tuned decisions (\eg, CFF vs.\ OP). Includes: determining the attack strategies, and adapt the automated discovery algorithms to the strategies.
Secondary target: teach people how hackers work, what they can do and why \softprot is important.
Secondary target: reassure experts that the engine works well
https://www.overleaf.com/project/5e1594c3b69e0300019fd9d9
\item 
automate the assessment of the attacks.\\
This requires the ability to estimate likelihood, and complexity (OWASP keywords). Other models of assessment (at least the fields) can be followed even if all the methods will be different.
Requires the ability to model what tools provide to attackers and when manual effort is needed.
May be hardcoded at the beginning, maybe more dynamic estimation can be achieved with more advanced automatic discovery algorithms

\item automatically identify the suitable {\softprot}s based the attack assessment

\item prioritize assets and {\softprot}s based on the attack assessment (may be superseded by full optimization)

\item automatically identify the {\softprot}s in synergy and compose them properly to maximize {\softprot} benefits

\item 
overhead estimation: evaluate the impact of the application of {\softprot}s in terms of performance and user experience 

\item estimate the impact of the {\softprot}s on assets/applications ({\softprot} level) without actually applying the {\softprot}s.
Needed to evaluate several combinations/{\softprot}s to choose the optimum.
Needed since assets and {\softprot}s are interleaved, it is not feasible to find a general model of prediction that is independent on the application to protect (universal metric, in netsec you just evaluate the fw and a few functionalities based on the service you want to protect)

\item
optimize the selection of the {\softprot}s to apply in order to mitigate the risks. Constraints: satisfy usability of the applications and the investments required 
Depends on previous items (impact without application and overhead estimation)
Do it before the end of the world (computational issues as for many optimization problems)

\item automatically mitigate the risks introduced by the mitigate phase itself (\ie asset hiding)
\item automatically apply {\softprot}s

\item the monitoring (as aspect of maintaining an up to date list of threats and risks and estimations and risk assessment procedures) depends on the economic model of the assets
\item 
\textcolor{red}{more?}
\end{itemize}
}
\fi

%% file: workflow.tex
\section{Proof-of-Concept Expert System for Software Protection}
\label{sec:workflow}

\label{sec:esp:kb}

Expert systems exist in cybersecurity since 1986 when Hoffman proposed one for the risk analysis of computer networks~\cite{hoffman1986risk}. From the initial intrusion detection systems~\cite{idesReq,audes} to  modern ones using AI~\cite{owensIDS}, expert systems have been used to automatically configure security controls~\cite{fwExpSys}, for post-incident network forensics~\cite{kimForensics,liaoForensics} and decision making.

The Expert System for Software Protection (\esp) is our \poc tool that implements a semi-automated \softprot risk analysis\footnote{In the ASPIRE project and in some cited papers, the ESP was called the ASPIRE Decision Support System (ADSS).}. Its complete code is available\footnote{\url{https://github.com/daniele-canavese/esp/}}, as well as a technical report on its inner workings~\cite{D5.11}, a user manual~\cite{D5.13}, and a demonstration video\footnote{\url{https://www.youtube.com/watch?v=pl9p5Nqsx_o}}.
The \esp is primarily implemented in Java as a set of Eclipse plug-ins with customized UI. 
It protects software written in C and needs source code access. 
The target users are software developers or \softprot consultants.
After the user manually annotates the assets in the source code, the \esp can generate what it considers optimally protected binaries and the corresponding security-server-side logic without human intervention.
As requested by the \softprot experts involved in the evaluation, a step-wise execution is also available where users can check and possibly override any information generated by the tool before executing the next step. 
\begin{figure}[t]
	\centering
    \includegraphics[width=6cm]{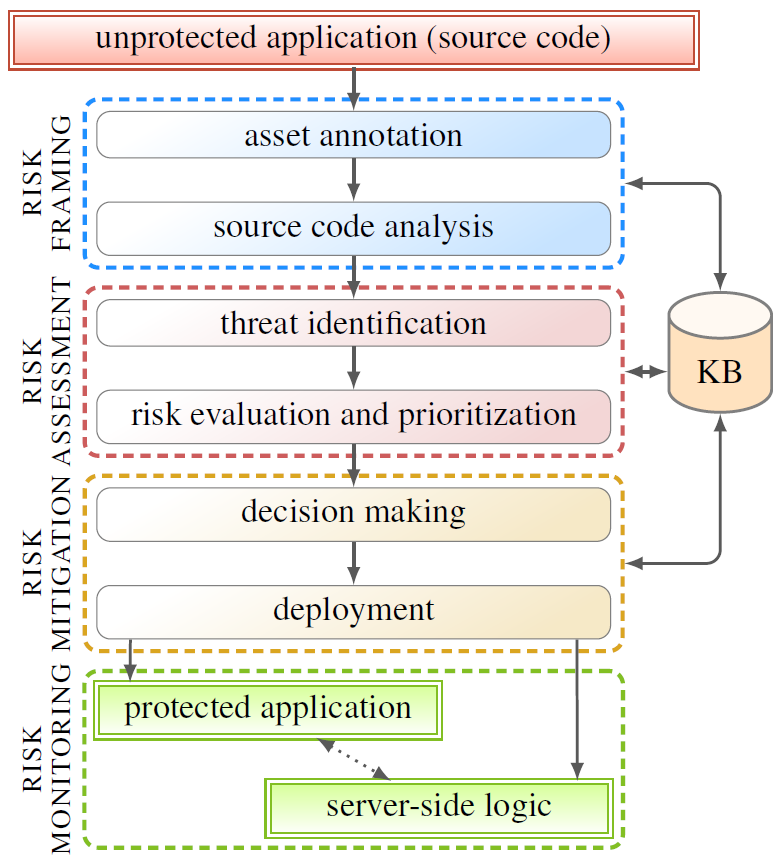}
	\caption{The \esp workflow.}
	\label{fig:workflow}
\end{figure}



Figure~\ref{fig:workflow} depicts the high-level workflow, split into the four phases discussed in Section~\ref{sec:requirements}. 

All the data needed for the risk analysis process, starting from the \emph{risk framing} information and including all the data obtained during the other three phases, are modeled and stored in a \kb.
The used model and corresponding \kb structure are designed specifically to support the reasoning methods needed for a software risk analysis~\cite{reganoMeta}.
For example, it allows representing context information, like the attacker model and the {\softprot}s available to mitigate the risks, and information about the application to protect, including the assets and abstract representations of the application code collected through code analyses.  More details about these constructs and the model are presented in Section~\ref{sec:esp:risk_framing}.

Next, the \esp performs the \emph{risk assessment} phase, whose details are provided in Section~\ref{sec:esp:risk_assessment}. This phase enriches the data in the model. %
It infers the possible attacks against the assets and assesses the risks against each asset by estimating the complexity of executing those attacks. 
The risk is evaluated by considering the software's structure and the attacker model, \ie the skills an attacker is likely to have, and asset values as defined by the user during the risk framing. 
The \esp's \emph{risk mitigation} phase, detailed in Section~\ref{sec:esp:risk_mitigation}, is also based on innovative methods. It uses \gls{ml} and optimization techniques to select the best \emph{solution}, i.e., the best sequence of {\softprot}s to be deployed and their configurations. It then automatically deploys it on the software to generate the protected application binaries.
If remote {\softprot}s are included in the selected solution, the deployment phase also generates the server-side logic to be executed on a trusted remote entity.

Finally, the \emph{risk monitoring} is performed. However, the \esp does not dynamically update the risk analysis process parameters. It only performs real-time integrity checking (as discussed in Section~\ref{sec:monitoring_relaesed}) depending on the methods implemented by the {\softprot}s used.

The \esp can also be used in two additional modes. It can be configured to propose a set of solutions that experts can manually edit to control the {\softprot} deployment fully. Moreover, it can be used to evaluate the effectiveness of solutions manually proposed by experts.

\subsection{Risk Framing in the ESP}
\label{sec:esp:risk_framing}

\begin{figure}[t]
	\centering
    \includegraphics[width=8cm]{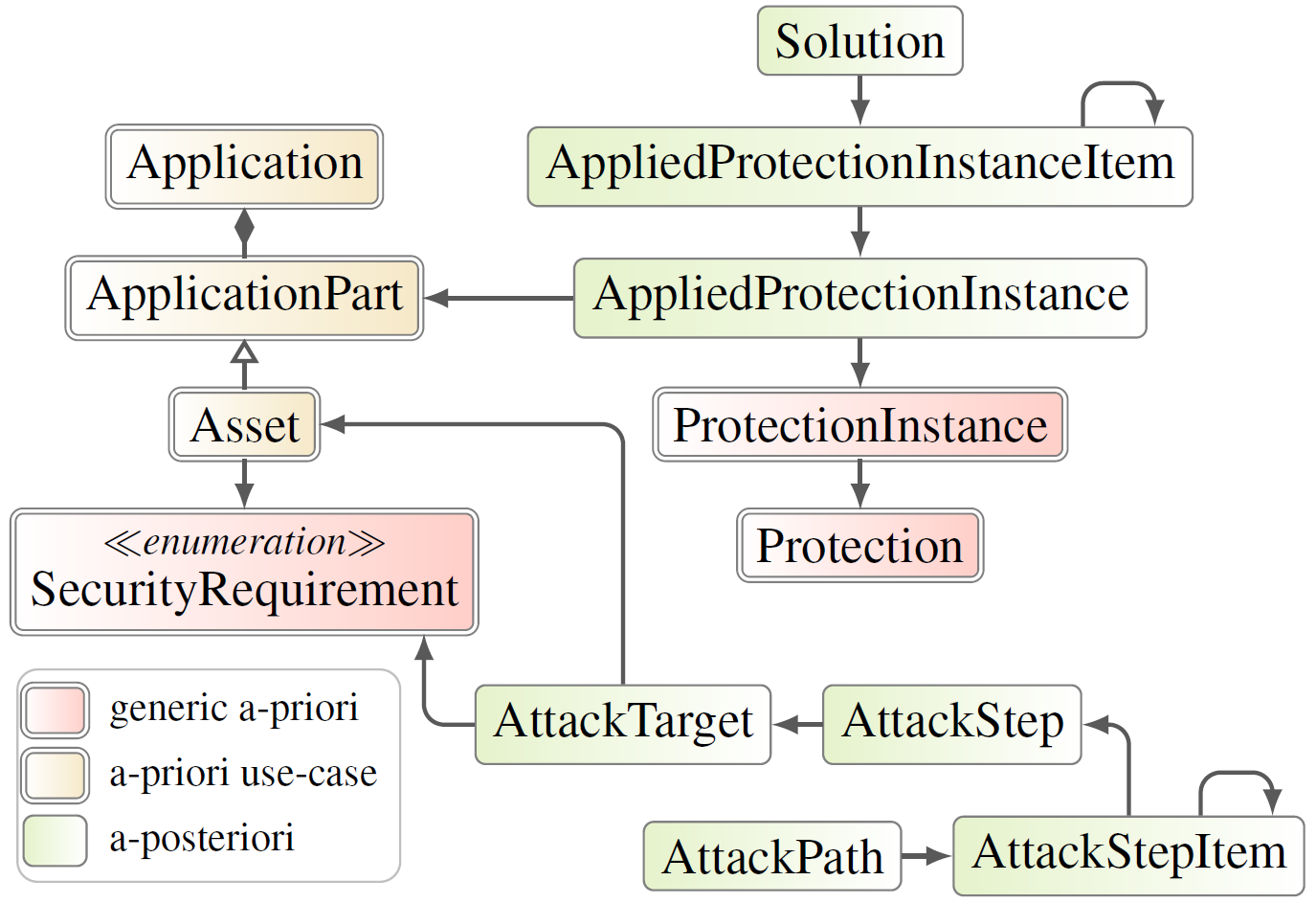}
	\caption{The top level of \esp meta-model to support modeling the relations between all relevant constructs.}
	\label{fig:metaModel}
\end{figure}

This tasks' purpose is to initialize all the constructs and their relations as needed for risk analysis, and to store them into a model formally defined in~\cite{reganoMeta} and named the \kb. It covers half of the models (\modellab{1}, \modellab{4}, \modellab{5}) highlighted in Section~\ref{sec:requirements}. Figure~\ref{fig:metaModel} presents the core classes, which will be discussed in the next sections. The \kb is instantiated as an OWL~2 ontology~\cite{owl2}.

The \emph{risk framing} starts with the preparation of the \kb with \emph{generic a-priori information}. This includes the core concepts and data not related to the specific application to be protected but relevant to framing the risk analysis process.
A priori information includes the assets types (\constructlab{1}, \constructlab{2}); the supported security requirements (\constructlab{8}, \constructlab{9}, \constructlab{12}, \constructlab{13},  \constructlab{14}); all the known attack steps and their characterization (\constructlab{4}, \constructlab{17}, \constructlab{18}, \constructlab{20}, \constructlab{21}, \constructlab{22}); the available {\softprot}s and their composability (\constructlab{24}, \constructlab{26}, \constructlab{27}, \constructlab{28}, \constructlab{29}); and the necessary constructs to evaluate risks and mitigations (\constructlab{23}, \constructlab{25}, \constructlab{30}, \constructlab{31}, \constructlab{32}, \constructlab{33}) that were discussed in Section~\ref{sec:framing}.
The user can also set preferences and analysis parameters (\constructlab{38}, \constructlab{44}), including hard and soft constraints and SDLC requirements (\constructlab{34}, \constructlab{36}), as well as the {\softprot}s to consider and the kinds of attacks to counter.

The \esp then performs a \emph{source code analysis} (\methodlab{2}) that populates the \kb with \emph{a-priori analysis-specific information} using the Eclipse C Development Toolkit\footnote{\url{https://projects.eclipse.org/projects/tools.cdt}}.
The analysis collects all the \emph{application parts}, \ie the variables, functions, and code regions. It determines additional information such as variables' data types and function signatures. It produces additional representations such as the call graph, which are useful for making decisions about the {\softprot}s to apply.

The \poc of the \esp supports confidentiality and integrity requirements. 
The user needs to annotate the source code with custom pragma and attribute annotations~\cite{D5.11,D5.13} to formally identify the code's assets and to specify their security requirements (\methodlab{1}, \methodlab{5}). 
The ESP then uses the call graph to identify potential secondary assets. These are listed in the GUI where the user can manually select which ones are to be considered assets in the later phases, and with which security requirements (\methodlab{3}, \methodlab{4}). Using only a call graph as a model (\modellab{2}) to find potential secondary assets that need to be manually confirmed is overly simple, and hence definitely a topic of future research.

Together with a-priori information, the \kb model represents \emph{a-posteriori information}, \ie data inferred and stored during later workflow phases such as the inferred attacks and the solutions.

{

In addition, the \esp offers a GUI to edit the framing information, e.g., to mark additional assets, characterize the attacker, and choose {\softprot}s. The GUI also allows importing and exporting risk framing data as XML or OWL files (\methodlab{6}).
This feature was appreciated during the validation as it allows augmenting the analysis with information that may be missed by the automatic but as of yet incomplete process, like the secondary assets that might be linked into a protected program as part of certain {\softprot}s.}


\subsection{Risk Assessment in the ESP}
\label{sec:esp:risk_assessment}

The risk assessment implements several methods to estimate the actual threats and risks. In the \emph{threat analysis}, the \esp uses backward reasoning methods (\methodlab{7}) to identify the attacks (\constructlab{39}) that can breach the primary assets' security requirements, and stores them in the \kb~\cite{reganoProlog}. 
This stage is roughly equivalent to the ISO27k ``identify risk'' step as discussed in sections~\ref{sec:standardizedRisk} and~\ref{sec:assessment}. 

The identified attacks are represented as a set (\modellab{6}) of \emph{attack paths} (\constructlab{3}). 
These are ordered sequences of atomic attacker tasks called \emph{attack steps} (\constructlab{4}).  
Attack paths are equivalent to attack graphs~\cite{attack_graphs} and can serve to simulate attacks with Petri Nets~\cite{petri_nets_attacks}. 
The attack steps that populate our \poc\ \kb originate from a study and taxonomy by Ceccato\etal~\cite{ceccatoTaxonomy,emse2019} and from data from industrial \softprot experts who participated in the ASPIRE project.

The attack paths are built via backward chaining (\methodlab{7}) as proposed in earlier work~\cite{basileOTP,reganoProlog} and implemented with SWI-Prolog~\cite{SWI-Prolog}.
An attack step can be executed if its premises are satisfied. It produces the results of its successful execution as conclusions.
The chaining starts with steps that allow reaching an attacker's final goal (the breach of a primary requirement) and stops at steps without any premise.
The search algorithm builds a proof tree with increasing depth and width, with exponential complexity. The \esp hence implements basic yet aggressive search space pruning to build an attack catalogue, e.g., by considering a maximum length for the inferred attack paths~\cite{ReganoPhd}. The proof tree models the actual threats (\modellab{6}). Its nodes can be seen as the exploited attack vectors (\constructlab{42}), of which the leaves form the identified attack surface (\constructlab{41}).

The \esp performs the \emph{threat impact evaluation} (\methodlab{8}) and \emph{risk prioritization} (\methodlab{9}) by assigning a \emph{risk index} (\constructlab{40}) to each identified attack path. Every attack step in the \kb is associated with multiple attributes, including the \emph{complexity} to mount it, the minimum \emph{skills} required, the availability of support \emph{tools} and their \emph{usability}.  
Additional attributes can be associated with entities trivially. 
Each attribute assumes a numeric value in a five-valued range.
For assessing the actual risks, the values of complexity metrics and software features (\constructlab{45}) computed on the involved assets (\methodlab{10}) with the available analysis tools (\constructlab{44}) are used as modifiers on the attributes (\constructlab{22}). 
For instance, an attack step labelled as medium complexity can be downgraded to lower complexity if the asset to compromise has a cyclomatic complexity below some threshold.

The risk index of an attack path is obtained by aggregating the modified attributes of its steps into a single value (\methodlab{8}). Our \poc is rather simple. Per attack step, it first aggregates all the step's modified attributes into a single attack step risk index. The attack path risk index is then computed by multiplying its steps' indices. Other aggregation functions are supported, such as summing the steps' indices, selecting maxima, and more complex features can easily be incorporated, like making the attack path risk index depend on how many different expert tools are required.

The report (\methodlab{13}) presenting the attack paths and the computed risk indices was welcomed by security experts (as will be discussed in more detail in Section~\ref{sec:esp:results}), amongst others because they serve as a starting point for evaluating the weaknesses of an application before more manual risk mitigation. Experts were interested in refining the identified, most risky paths into more concrete sequences of attack operations, and in some cases, they would have manually updated the risk indices.
In our \poc, the attack steps are coarse-grained, such as ``locate the variable using dynamic analysis'' and ``modify the variable statically''.
This is an important limitation. As Section~\ref{sec:identification_threats} discusses, understanding how much refinement is needed is an open research question.

\subsection{Risk Mitigation in the ESP}
\label{sec:esp:risk_mitigation}

Before presenting the ESP's risk mitigation process \methodlab{16}, we introduce more precise constructs. 
In the ESP, a \emph{\softprot} is a specific implementation of a \softprot technique by a specific \softprot tool.
For instance, control flow flattening~\cite{wangFlatteningTechReport} as applied by Diablo in the \actc and by Tigress are considered distinct {\softprot}s~\cite{tigress,diablo}.\footnote{\url{https://github.com/aspire-fp7/actc} and \url{https://tigress.wtf/}}
A \emph{protection instance} (PI) is a concrete configuration of a \softprot technique.
The \esp can use the PI to drive the \softprot tool to apply a {\softprot} technique on a chosen application part.
Depending on the available parameters, multiple PIs can be defined for the same {\softprot}. 
%
An \emph{applied PI} is the association of a PI with an application part, which states that the PI has been selected to be applied to the part. 
A \emph{candidate solution} is a sequence of applied PIs. It is ordered because of composability and layering requirements and benefits (\constructlab{27}).

The \esp first searches for \emph{suitable {\softprot}s}. These are {\softprot}s that impact attributes of the listed attack steps (\methodlab{19}). For example, they are able to defer an attack step.
Each PI is associated with a formula that alters these attributes for each attack step. 
After the application of a {\softprot}, the risk index of the attack steps and paths are re-assessed. 

The formulas also consider complexity metrics (\constructlab{25}) computed on the protected assets' code. This way, the ESP incorporates Collberg's prescription of \emph{potency}~\cite{collberg1997taxonomy} (\constructlab{30}) as a measure of the additional effort that attackers have to invest on protected code. 
The parameters to be used in the formulas for evaluating the impact of SPs on attack steps are stored in the \kb. 
They are based on a survey among the developers of all {\softprot}s integrated into the ASPIRE \softprot tool flow~\cite{D5.11}, whom we asked to score the impact of their {\softprot}s on a range of attack activities in terms of concrete impacts. 
These include the impact on human comprehension difficulty by increasing code complexity, the impact of moving relevant code fragments from the client-side software to a secure server not under the control of an attacker~\cite{renewability,codeMobility,viticchie2016reactive}, the impact on the difficulty of tampering through anti-tampering techniques with different reaction mechanisms and monitoring capabilities~\cite{viticchie2016reactive}, and the impact of preventive {\softprot}s such as anti-debugging~\cite{circulardebugging,diabloSelfDebugging}. 
The survey results were complemented with expert feedback and validated in pen test experiments~\cite{ceccatoTaxonomy,emse2019}.

Additional modifiers are activated when specific combinations of PIs are applied on the same application part. They model the impact of layered {\softprot}s (\constructlab{28}) when recomputing the risk indices and synergies between {\softprot}s. The existence of synergies (\constructlab{29}) was part of the mentioned survey. 

Candidate solutions must also meet cost and overhead constraints (\constructlab{33}). 
Our \poc filters candidate {\softprot}s using five overhead criteria: client and server execution time overheads, client and server memory overheads, and network traffic overhead. 

Finally, the \emph{{\softprot} index} associated with a candidate solution is computed based on the recomputed risk indices of all discovered attack paths against all assets, weighted by the importance associated with each asset. The {\softprot} index is the ESP's instantiation of residual risk (\constructlab{47}).

\subsubsection{Asset Protection Optimization}
\label{sec:esp:risk_mitigation:optimization}
\label{sec:esp_optimization_approach}
The \esp finds the mitigations by building an optimization model that it solves with a game-theoretic approach (\methodlab{23}). The \esp tries to combine the suitable {\softprot}s to build the optimal layered  solutions, i.e., the candidate solution that maximizes the {\softprot} index and satisfies the constraints.

Computing the {\softprot} index by re-computing the risk index requires knowledge of the metrics on the protected application. As applying all candidate solutions would consume an infeasible amount of resources, we have built an \gls{ml} model to estimate the metrics delta after applying specific solutions without building the protected application~\cite{reganoMetric}. 
The ESP's ML model has been demonstrated to be accurate for predicting variations of up to three PIs applied on a single application part. With more {\softprot}s, however, the accuracy starts decreasing significantly. This issue seems to be solvable with larger data sets and more advanced \gls{ml} techniques.

The \esp uses the same predictors to estimate the overheads associated with candidate solutions. Per PI and kind of overhead, the \kb stores a formula for estimating the overhead based on complexity metrics computed on the vanilla application. These formulas were determined by the developers of the different {\softprot}s integrated into the tool flow of the ASPIRE project. 

Combinations greatly increase the solution space. To explore it efficiently and to find (close to) optimal solutions in an acceptable time, the \esp uses a game-theoretic approach, simulating a non-interactive \softprot game (\methodlab{24}). 
In the game, the defender makes one first move, \ie proposes a candidate solution for the protection of all the assets. Each proposed solution yields a base {\softprot} index, with a positive delta over the risk index of the vanilla application. 
Then the attacker makes a series of moves that correspond to investments of an imaginary unit of effort in one attack path, which the attacker selects from the paths found in the attack discovery phase. Similarly to how potency-related formulas of the applied {\softprot}s yield a positive delta in the {\softprot} index, we use resilience-related formulas that estimate the extent to which invested attack efforts eat away parts of the {\softprot} potency, thus decreasing the {\softprot} index. These formulas are also based on expert feedback. We refer to Regano's thesis for more details on this game-theoretic approach that uses mini-max trees and a number of heuristics to yield acceptable outcomes in acceptable times~\cite{ReganoPhd}. 

After solving the game, the \esp shows the best {\softprot} solutions (\constructlab{48}, \constructlab{49}) it found, i.e., the best first moves by the defender, from which the user can choose one. The \esp then invokes the automated \softprot tools to apply the solution, as will be explained in Section~\ref{sec:esp:workflow:solDep}. 

\subsubsection{Asset Hiding}
\label{sec:esp:risk_mitigation:hiding}
\label{sec:esp:workflow:hiding}
As discussed in Section~\ref{sec:framing}, {\softprot}s are not completely stealthy because they leave fingerprints. 
In a previous paper \cite{reganoL2P}, we proposed a solution to this problem based on the refinement of existing {\softprot} solutions with additional {\softprot}s also deployed on non-asset code regions (\methodlab{22}). Those lure the attacker into analyzing such regions in lieu of the assets' code, thus hiding the assets from plain sight. We have devised three asset-hiding strategies. In \emph{fingerprint replication}, {\softprot}s already deployed on assets are also applied to other code parts to replicate the fingerprints such that attackers analyse more parts. With \emph{fingerprint enlargement}, we enlarge the assets' code regions to which the {\softprot}s are deployed to include adjacent regions such that attackers need to process more code per region. With \emph{fingerprint shadowing}, additional {\softprot}s are applied on assets to conceal fingerprints of the chosen {\softprot}s to prevent leaking information on the security requirements.

The \poc \esp hides the protected assets in an additional decision making step. 
In this step, it adds \emph{confusion indices} to the {\softprot} indices, which are computed by an ad hoc formula built to estimate the additional time the attacker needs to find the assets in the application binary after the application of hiding strategies. The computation of the confusion indices requires estimating the code complexity metrics after the application of the {\softprot}s.
To build this model, we have studied the effects of the hiding strategies for the {\softprot}s devised during the ASPIRE project. The results of this study, stored in the \esp \kb, are used to compute the confusion index.

Starting from the solutions generated via the game-theoretic approach, the \esp proposes additional application parts to protect by solving a Mixed Integer-Linear Programming (MILP) problem, expressed as a heavily customized instance of the well-known 0-1 knapsack problem~\cite{knapsack-book} that maximizes the confusion index and uses overhead as weight in constraints.
%
The MILP problem is solved using an external solver; the \poc\ \esp supports lp\_solve and IBM CPLEX Optimizer\footnote{See \url{http://lpsolve.sourceforge.net/5.5/} and \url{https://www.ibm.com/analytics/cplex-optimizer}.}. 

In between the asset protection optimization and the asset hiding, no measurement is done on code on which {\softprot}s are deployed. The ESP's decision making is a single-pass process (\methodlab{20}).

\subsubsection{Deployment}
\label{sec:esp:risk_mitigation:deployment}
\label{sec:esp:workflow:solDep}

The final step in the \esp workflow deploys the solution on the target application (\methodlab{17},\methodlab{26}). The solution is chosen by the user amongst the ones presented by the \esp. The result of this step (and of the whole workflow) is the protected binary plus source code for the server-side components for selected online {\softprot}s. 
The \esp deploys a solution by driving automatic {\softprot} tools. At the time of writing, the \esp supports Tigress, a source code obfuscator developed at the University of Arizona, and the \actc, which automates the deployment of {\softprot} techniques developed in the ASPIRE FP-7 project~\cite{D5.11,D5.13}. Table~\ref{tab:protections} summarizes the {\softprot} techniques supported by the \esp.

\begin{table}
    \centering
    {\small
    \begin{tabular}{lcccc}
        \toprule
        \multirow{2}{*}{\textsc{protection type}} & \multicolumn{2}{c}{\textsc{requirements}} & \multicolumn{2}{c}{\textsc{tool}} \\ 
        \cmidrule(lr){2-3} \cmidrule(lr){4-5}
        & \textsc{confidentiality} & \textsc{integrity} & \textsc{ACTC} & \textsc{Tigress}\\
        \midrule
            anti-debugging             & \faCheckCircle & \faCheckCircle & \faCheckCircle & \faCircleO\\
            branch functions           & \faCheckCircle & \faCircleO     & \faCheckCircle & \faCircleO\\
            call stack checks          & \faCircleO     & \faCheckCircle & \faCheckCircle & \faCircleO\\
            code mobility              & \faCheckCircle & \faCheckCircle & \faCheckCircle & \faCircleO\\
            code virtualization        & \faCheckCircle & \faCheckCircle &\, {\faCircleO}\tablefootnote{The ACTC provides limited support for code virtualization, meaning that it is not reliably applicable to all code fragments. Hence the ESP does not consider it a potential protection instance.} & \faCheckCircle\\
            control flow flattening    & \faCheckCircle & \faCircleO     & \faCheckCircle & \faCheckCircle\\
            data obfuscation           & \faCheckCircle & \faCircleO     & \faCheckCircle & \faCheckCircle\\
            opaque predicates          & \faCheckCircle & \faCircleO     & \faCheckCircle & \faCheckCircle\\
            remote attestation         & \faCircleO     & \faCheckCircle & \faCheckCircle & \faCircleO\\
            white-box crypto     & \faCheckCircle & \faCheckCircle & \faCheckCircle & \faCircleO\\
        \bottomrule
    \end{tabular}
    }
    \caption{SPs supported by the \esp, with enforced security requirements and tools used to deploy the SPs. For each tool, we only mark techniques supported on our target platforms, i.e., Android and Linux on ARMv7 processors.}
    \label{tab:protections}
\end{table}

Finally, we point out that the \esp has been engineered to be extensible. All the modules can be replaced with alternative components. For example, the risk assessment based on backward reasoning could be replaced with a more advanced attack discovery tool, the only constraint being {that it needs to produce} output compliant with the {\softprot} meta-model. 
It is also possible to support new {\softprot}s. 
It is enough to add the required information into the \kb, such as the evaluation of strengths and impacts on attack steps, conflicts, and synergies with other {\softprot}s plus all parameters of the discussed formula. The only demanding activities are training the \gls{ml} algorithms to predict how new {\softprot}s alter the metrics, and the automation of the deployment of the {\softprot}s.

\subsection{Risk Monitoring in the ESP}
\label{sec:esp:risk_monitoring}

If the selected {\softprot}s include online {\softprot}s such as code mobility~\cite{codeMobility} and reactive remote attestation~\cite{viticchie2016reactive}, the \esp generates all the server-side logic, including the backends that perform the risk monitoring of the released application. This includes the untampered execution as checked with remote attestation, but also the communication with the code mobility server (\methodlab{30},\methodlab{31}).

Our \poc does not automatically include the feedback and other monitoring data such as the number and frequency of detected attacks and compromised applications, and server-side performance issues.
The knowledge base needs to be manually updated using GUIs to change risk framing data related to attack exposure and {\softprot} effectiveness. Issues related to insufficient server resources also need to be addressed independently; the \esp only provides the logic, not the server configurations. 



\subsection{Coverage}
\label{sec:coverage}


As could already be seen in Tables~\ref{tab:constructs},~\ref{tab:models}, and~\ref{tab:methods}, the ESP covers many of the constructs, models, and methods we positioned in the overall risk management approach in Section~\ref{sec:requirements}.  Albeit to some extent in a rudimentary form, as can be expected from a proof-of-concept tool, the ESP instantiates 36 of the 50 identified constructs, 5 of the 6 discussed models, and 21 of the 32 methods.

All of the instantiated artifacts were required to meet the objectives and requirements of the ASPIRE project. 
The reasons why the other 14+1+11 artifacts are not instantiated in the ESP are that ASPIRE research project plan was drafted and executed before the development of our vision on standardization, and that the project had a limited time frame and resources, and hence a limited scope and set of requirements to meet.

\changed{The non-instantiated artifacts relate to five major limitations that in our opinion do not impact the possibility to build an entire \softprot workflow that includes mostly automated tasks. Indeed, all the activities that we have not (yet) automated can be performed manually, so if a fully automated approach would be proven impossible at some point in the future, a semi-automated approach is certainly possible.}

First, ASPIRE focused solely on the technical threats, neglecting the relationship with business risks. Constructs \constructlab{6} and \constructlab{11}, model \modellab{3}, and methods \methodlab{14}, \methodlab{15}, and \methodlab{28} that relate to business models were hence out of scope. With the attack exploitation phase (\constructlab{11}) being out of scope, the decision support tool did not have to differentiate between threats in that phase and in the attack identification phase, and hence no tool support to treat (\constructlab{10}) as an explicit construct was needed.
\changed{Existing methods to link technical threats and constraints to business risks are available as discussed in Section~\ref{sec:req:prioritization}, if not automated then certainly relying on human judgments.
Furthermore, providing support for considering multiple attack phases requires no fundamental changes to the used models and methods.}

Second, another scope limitation was that ASPIRE only considered the protections of application instances in isolation, not as they evolve over time, and with the \softprot tool having white-box access to all relevant application code. Hence constructs \constructlab{7}, \constructlab{37} and \constructlab{46}, and methods \methodlab{11}, \methodlab{12}, \methodlab{27}, and \methodlab{29} were out of scope. 
\changed{Experts can manually deal with those \sdlc issues in case future research would fail to provide automated solutions.}

Third, whenever functional requirements were stated, such as the need to deploy a copy-protection scheme, only one implementation of that functionality was developed. Hence no decisions needed to be made on how to meet those requirements, making decision support for functional requirements (\constructlab{13}) irrelevant within the project. 
\changed{In general, decision support for functional requirements is simpler than for non-functional requirements: functional requirements are typically expressed as ``some form of protection functionality X needs to be included.'' If anything, such  requirements limit the search space that the \softprot optimization algorithms need to explore, rather than complicating it.}

Fourth, whereas Section~\ref{sec:motivation_formalization_automation} argued for maximal formalization and automation to minimize the potential reduction in precision that can stem from the subjective expert judgments, within the ASPIRE project complete automation was not considered viable yet. The involvement of experts in making judgments was still accepted, so some aspects were not formalized and automated but instead left to human experts. This is the case for methods \methodlab{18} and \methodlab{132} for validating that the \softprot deployment is in line with made choices and requirements; for constructs \constructlab{5}, \constructlab{15}, and \constructlab{16} that serve to identify which application parts require protection despite not being primary assets; for identifying the path of least resistance (\constructlab{43}) among the enumerated attack paths; and for iterative mitigation decision making (\constructlab{50}, \methodlab{21}). An expert can use the \esp manually in an iterative manner, but the \esp does not automate this. Finally, while the \softprot tool developed in ASPIRE considers profile information for minimizing the performance impact of injected control flow obfuscations, the \esp does not consider profile information (\constructlab{35}) for selecting {\softprot}s. It is hence left up to the human expert to manually exclude expensive {\softprot}s for assets on which they cannot be afforded. 

Fifth, the one remaining unsupported artifact is that of cookbooks with \softprot recipes (\methodlab{25}). Those cookbooks are only intended as backups for when automated \softprot selection is not supported. They are hence superfluous in the \esp.


\section{Evaluation of the Instantiated Artifact}
\label{sec:esp:results}
\label{sec:evaluation}
The \esp has been designed and implemented to answer RQ1 \changed{(whether automated decision support tools can assist experts with the deployment of SPs and
the use of SP tools)} within the scope and requirements of the ASPIRE project.
However, this evaluation also provides evidence to answer RQ3 \changed{on which parts of a standardized risk management approach to \softprot can already be automated}, as it provides a lower bound on the set of those parts. Moreover, since the \esp implements a NIST-based four-phase risk analysis approach, a positive evaluation of the tool provides evidence that \changed{modelling {\softprot} as a risk management task is feasible. The \esp hence partially backs up the list of constructs, models, and methods discussed in Section~\ref{sec:requirements} to answer RQ2 on which artifacts a standardized risk management approach in the domain of \softprot needs to entail.}

For this evaluation, the research question RQ1 is further split into more precise technical questions according to ISO/IEC 9126-1:2001 evaluation criteria. \changed{Table~\ref{table:RQs} lists these questions and the main results/answers.} We answer RQ1.a--RQ1.c with the qualitative evaluation in Section~\ref{sec:qualitative-evaluation:results} of which the design is first discussed in Section~\ref{sec:design_qualitative_eval}. RQ1.d is answered in Section~\ref{sec:quantitative-assessment} and Section~\ref{sec:time}. 

\begin{table}
\centering
{\small
\changed{
\begin{tabular}{l p{11cm}}
\multicolumn{2}{l}{\textbf{Usability}}\\
\hline
RQ1.a.1 & Is the \esp usable by software protection experts?\\
& $\rightarrow$ \hangindent = 0.4 cm positive answer: experts did not have problems in using the tool and understanding the meaning of all the artifacts \\
RQ1.a.2 & Can the \esp become part of the experts' daily workflow?\\
& $\rightarrow$ \hangindent = 0.4 cm positive answer: experts reported that the \esp was suitable for their daily job but further effort is needed for a better business alignment\\
RQ1.a.3 & Could the \esp be used also by software developers with limited or no background in software protection?\\
& $\rightarrow$ \hangindent = 0.4 cm partially positive answer: proper tuning of the \esp requires a solid background in \softprot, pro\-tection will be less effective when using the \esp as a push-one-button tool\\

\\
\multicolumn{2}{l}{\textbf{Correctness}}\\
\hline
RQ1.b & Does the \esp propose appropriate combinations of protections to protect the assets and to hide them properly?\\
& $\rightarrow$ \hangindent = 0.4 cm positive answer: experts analysed all artifacts produced by the \esp, compared them to data from tiger teams, and judged them as correct or appropriate\\
\\
\multicolumn{2}{l}{\textbf{Comprehensibility and Acceptability}}\\
\hline
RQ1.c.1 & Is this output useful to comprehend and assess the tasks (automatically) performed by the \esp?\\ 
& $\rightarrow$ \hangindent = 0.4 cm positive answer \\
RQ1.c.2 & Is the output produced by the \esp useful to help software protection experts manually perform their job?\\
& $\rightarrow$ \hangindent = 0.4 cm answer is limited: the inferred attacks has been judged too coarse-grained, this affects the precision in automatically deciding the mitigating protections  \\
\\
\multicolumn{2}{l}{\textbf{Efficiency}}\\
\hline
RQ1.d.1 & Is the \esp fast enough to produce valid solutions in a useful time? \\
& $\rightarrow$ \hangindent = 0.4 cm positive answer\\
RQ1.d.2 & Is the complexity of the algorithms it uses acceptable for the size of the tasks it has to perform?\\
& $\rightarrow$ \hangindent = 0.4 cm positive answer: experts and developers considered the execution times acceptable, scalability may be an issue due to worst-case complexity, but heuristics and optimization allowed producing solution in useful time\\
\end{tabular}
}
}
\caption{Refinement of RQ1 into concrete \esp research questions and this paper's main answers to them.}
\label{table:RQs}
\end{table}

\subsection{Design of the Qualitative Evaluation}

\label{sec:design_qualitative_eval}

The \esp has been validated with expert {\softprot} users drawn from the ASPIRE project consortium and advisory boards. This qualitative evaluation is a snapshot of experts' opinions about the final ASPIRE PoC at the end of the project in Q4 2016, when they were last available to us.

In ASPIRE, each industrial partner provided an Android app use case: a One-Time Password generator for home banking apps, an app licensing scheme, and a video player with \drm for protected content.
Each app included security-sensitive code and data elements in dynamically linked, native libraries written in C. 
Those libraries served as reference use cases for all research. 
They were designed and implemented to represent the industrial partners' commercial software. 
Being sensitive, the industrial partners only gave access rights to their use cases to academic partners, not to each other. Less sensitive information on the use cases, their assets, their security requirements, the software features obtained with the analysis tools, and their experts' assessments are available in a public report that presents a joint validation of all project results~\cite{D1.06}. 
Table~\ref{tab:useCase} presents the \sloc metrics and the number of assets. Clearly, the \esp was not evaluated merely on toy examples.

The evaluation of the ESP in ASPIRE was planned to be done by two experts per industrial partner: one \emph{internal expert} familiar with the project and involved in it, and one \emph{external expert} that was not involved in the project. However, one of the three partners only made the internal expert available, who hence participated in both roles. According to \feds, this limitation in the number of experts  and the limitation in the scope (each company's experts evaluated the artifact using only their own use case) needs to be considered a significant constraint.

\begin{figure}[t]
\centering
\includegraphics[width=.97\textwidth]{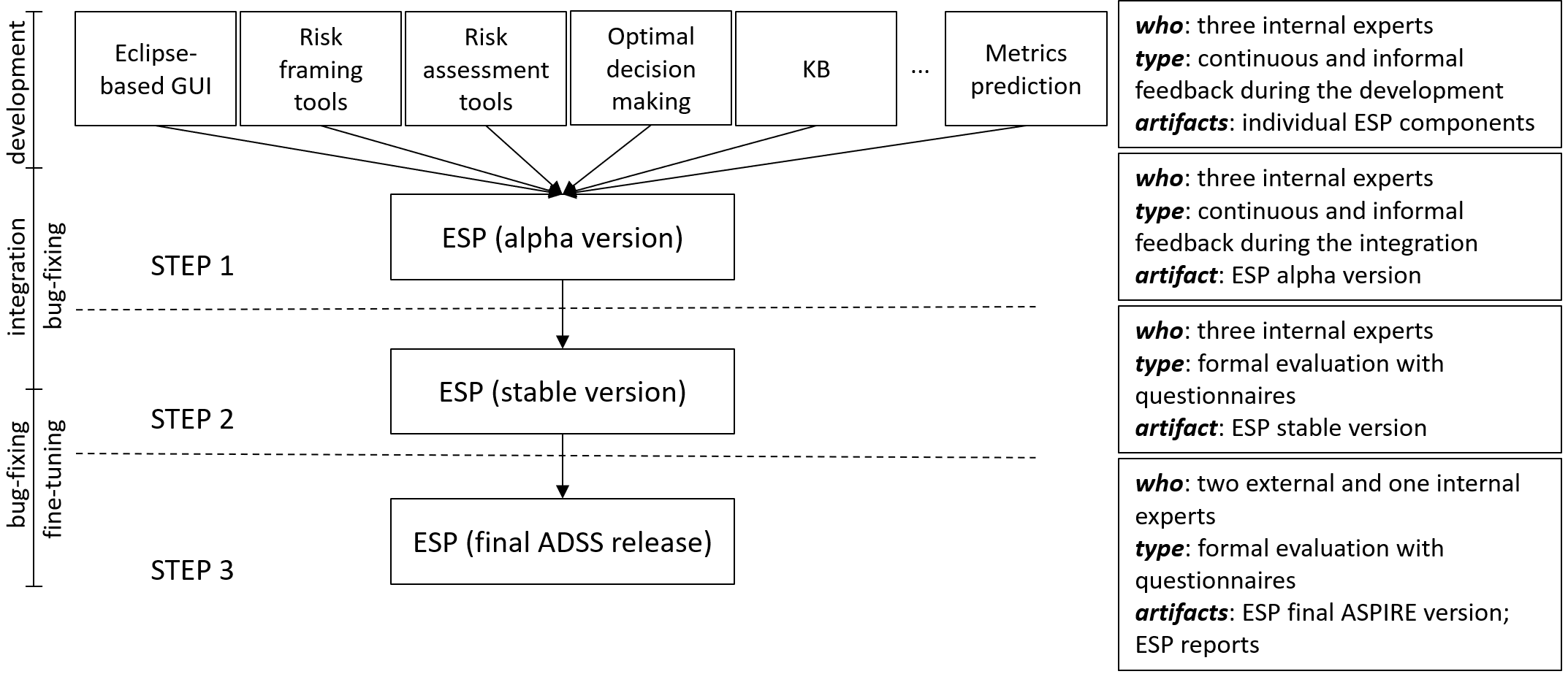}
\caption{Graphical representation of the qualitative evaluation process.}
\label{fig:evaluation_steps}
\end{figure}

We then organized the qualitative evaluation of usability, correctness, comprehensibility, and acceptability in three steps, as visualised in Figure~\ref{fig:evaluation_steps}.

\subsubsection{Step 1: Early Internal Expert Assessment}
\label{sec:early_assessment}
\changed{During the ESP's development} the internal experts performed a qualitative analysis of the prototype artifact to improve early versions. 
They followed the design and development of the \esp throughout the three-year project and continuously monitored the results of the {\softprot} on their use cases, using the \esp to provide \changed{constructive} feedback. 
They analysed methods, models, constructs, and instantiations of the individual \esp components to check the correctness of their results. They were also involved in designing the workflows to comply with their corporate needs.
The result was an alpha version of the tool components integrated into the workflow. 

\changed{This alpha version was then evaluated as a whole by the internal experts.} 
Each internal expert was asked to protect their software only using the artifact with the support of the \esp developers. They manually annotated the assets in the source code using the \esp GUI and then used the \esp to identify the best {\softprot}s and their configuration parameters. 
They then discussed and analysed the identified threats and the selected {\softprot}s, as well as the entire decision making process, on which they then commented in detail. We collected their inputs through interviews during face-to-face meetings, ad-hoc calls, and emails. Since these inputs were collected during the normal project development, the collection was managed informally. The experts also gave a qualitative assessment of the correctness of the artifact's used models.

Moreover, the effectiveness of the ESP's selection of {\softprot}s was tested against the judgment of other experts. 
Indeed, the use cases' developers and security architects proposed the best combination of {\softprot}s for each of the assets, according to their expertise and experience.
Each of the three use cases protected with the best combination were then pen tested by two external pen testers per use case for several weeks to establish the attack paths. 
The pen testers reported on the attacks that were prevented entirely for the DemoPlayer use case within the pen test time frame, and that the attacks were delayed effectively for the other two use cases\footnote{This material is available in Section~5 of the public ASPIRE Validation Report~\cite{D1.06} and in sections 8--11 of the public ASPIRE Security Evaluation Methodology Report~\cite{D4.06}}.

\subsubsection{Step 2: Final Internal Expert Assessment}
\changed{Towards the end of the ASPIRE project, a first stable version of the whole \esp was available. On the basis of this version, experts were asked to assess the \esp by answering a set of open-ended questions, which are provided in~\ref{sec:appendix}.
The experts' answers were used to develop the final release of the tool during the ASPIRE project. Their answers are not reported in this paper as they were considered confidential material. Several calls took place with those experts to clarify questions and to ensure that we interpreted their answers correctly.}

This assessment of the design and implementation of the \esp constitutes a qualitative evaluation \changed{in a naturalistic scenario, which, using FEDS terminology, means that a real system (artifacts) is used by real users to solve real problems~\cite{Prat-taxonomy}. In our case, the artifact consists of the first complete version of the \esp; the real users are the industrial experts; and the real problem is the selection of protections to mitigate real \mate attacks (as evaluated in pen tests by pen testing experts~\cite{emse2019}) on applications developed by the industrial project partners to be equivalent to their commercial applications, i.e., feature the same types of assets, similar functionality, and similar complexity.}  


\subsubsection{Step 3: Assessment with External Experts}
\label{sec:external_expert_design}
\changed{Finally, external experts from the industrial partners, which had no prior insights or bias regarding the \esp, were involved in a qualitative evaluation of the final \esp version in the ASPIRE project.} We used the same questionnaire for this evaluation.

We prepared a virtual machine (VM) with a running copy of the \esp pre-configured with all the manual operations already performed by their colleagues. The assets were already annotated, and the other \esp running parameters were set to default values, which they were allowed to modify. 
However, they experienced configuration issues when integrating the \esp VM with the {\softprot} tools. 
To make good use of their extremely limited time, we therefore also executed the tool with the pre-configured information and all automation enabled and provided them with the output generated by the tool in the form of a report\footnote{The reports are available at \url{https://github.com/daniele-canavese/esp/tree/master/reports}. The ESP user manual~\cite{D5.13} describes how to interpret the different parts of those reports. In two of the three reports, we renamed identifiers of code and data elements (such as function names) in consideration of the two companies' confidentiality requirements. Apart from that, the linked reports are identical to the ones assessed by the experts. A summary of the most significant data, their interpretation, and the major findings is presented on the aforementioned GitHub site.}.
The expert was then asked to assess the identified threats, the selected {\softprot}s, and the selected properties of the evaluation.

\subsection{Qualitative Evaluation Results and Discussion}
\label{sec:qualitative-evaluation:results}


Overall, based on the analysis of their questionnaires, the experts have judged the \esp as promising and potentially effective because of the high level of automation and configurability (including the possibility to override default configurations) and the detailed output.
Nonetheless, they were skeptical about extending the tool's use to software developers with a more limited background in {\softprot} as this background is needed for understanding the artifacts, making decisions, and evaluating results. This is not preventing software developers from using the ESP as a push-of-the-button tool and having their applications protected. However, they feared that in the push-of-the-button mode, the applications risked being less protected than under the supervision of experts. 
Furthermore, the acceptability showed limitations at the level of integration with their daily work and tool chains, which means that further effort is needed to ensure so-called alignment with business~\cite{Prat-taxonomy}.
The usability was hence assessed positively for experts in {\softprot} (RQ1.a.1), positively with limitations for the integration into the experts' tool flow (RQ1.a.2), and only partial for developers (RQ1.a.3).

Among the data extracted by the tool, experts highlighted the importance of making decisions by considering the application structure and metrics because results are to be tailored to the target application.
They also appreciated that all the data extracted and represented in the \kb are structured using a formal meta-model, as this reassured them of the correctness of the inferences.

\changed{Experts analysed the attack paths inferred by the tool as well as the {\softprot}s solutions that were proposed by the optimization process to mitigate the inferred attacks. The experts compared those solutions} to the ones they (or their colleagues) had assembled manually earlier on during the project as part of the requirements formulation.
Solutions have been validated in terms of achieved security for the assets, preservation of the application business logic, and containment of the inevitable slow-down of the protected application \wrt the original one. 
Furthermore, the attack paths have been compared with the real attacks discovered by the professional pen testers previously involved, as discussed in Section~\ref{sec:early_assessment}. In particular, the inferred solutions have been judged as appropriate to protect the use case code and effective in blocking the inferred attack paths and the real attacks reported by the pen testers. In addition, the protected binaries were evaluated as semantically unaltered and usable: they still delivered the original observable IO-relation without excessive overhead introduced by the {\softprot}s.
It is worth mentioning that, as is the case for all the risk analysis processes, there is not a correct set of answers forming a ground truth. The experts hence provided their qualitative estimations of solution effectiveness.

The main flaw of \esp reported by the experts is that inferred attack steps were too coarse-grained because of too generic attack rules. This limitation has a technical impact on the possibility of making fine-grained decisions on the {\softprot}s to use. For instance, consider the listed attack step \lstinline|staticallyLocate('ProvisioningManager_LaunchProcess.r16'(attacker))|. This denotes an attacker disassembling the binaries with static tools such as Radare2 or IDA Pro to locate basic block \texttt{r16} in function \texttt{LaunchProcess}. From this, it is possible to infer that obfuscation is needed. However, on the basis of this information alone, it is impossible to determine which specific obfuscation technique should be preferred without looking at the actual code. So expert judgment and interaction to refine the \softprot selection by the \esp is currently still required. It is definitely possible to populate the \kb with more fine-grained attacks steps, but as already mentioned in Section~\ref{sec:identification_threats}, further research is needed to determine the best level of granularity to model and enumerate attacks steps.

From all of the above, we conclude a positive evaluation of the correctness of the artifact (RQ1.b) and the comprehensibility of the artifact-generated data (RQ1.c.1). The usability of specific artifacts has limitations (RQ1.c.2).

The efficiency has been measured with a quantitative assessment, as will be discussed in Section~\ref{sec:quantitative-assessment}. In addition, no experts reported issues with the performance of the artifact.

Overall, the evaluation result is hence positive. 
Quoting from the related project deliverable, \textit{``after the analysis of the validation data, the experts concluded that the tool has a very high potential''} to be used in their everyday tasks and to enter their current workflow in the near future, even if some had doubts on the maturity of the tool and its readiness to be used to protect commercial applications with all of their SDLC intricacies and complexity. For an artifact developed as a research proof-of-concept, this should of course not come as a surprise. 

We conclude that a large part of the proposed risk management approach can indeed be automated through decision support tools, as identified by many of the checkmarks in Table~\ref{tab:methods} that provide an answer to RQ3. While not yet capable of completely replacing human experts, those proof-of-concept automated tools have shown promise to aid users of \softprot tools.

\begin{table*}[t]
	\centering
	{\small
	\begin{tabular}{lrrrrrr}
		\toprule
		\multirow{2}{*}{\textsc{application}} & \multicolumn{2}{c}{\textsc{C}} & \multirow{2}{*}{\textsc{Java}} & \multirow{2}{*}{\textsc{C++}} & \multirow{2}{*}{\textsc{assets}}\\
		\cmidrule(lr){2-3}
		& \textsc{sources} & \textsc{headers} & & &\\
		\midrule
		DemoPlayer & 2,595 & 644 & 1,859 & 1,389 & 25 \\
		LicenseManager & 53,065 & 6,748 & 819 & 0 & 43 \\
		OTP & 284,319 & 44,152 & 7,892 & 2,694 &  25 \\
		\bottomrule
	\end{tabular}
	}
	\caption{Lines of source code counts of the ASPIRE validation use cases.}
	\label{tab:useCase}
\end{table*}

\subsection{Technical Assessment}
\label{sec:quantitative-assessment}
\label{sec:technical-assessment}

\begin{table}[t]
	\centering
	{\small
	\begin{tabular}{lcccc}
		\toprule
		\textsc{application} & \textsc{Total} & \textsc{Framing} & \textsc{Assessment} & \textsc{Mitigation} \\
		\midrule
		DemoPlayer & 145.6 & 0.1 & \ 76.3 & \ 69.1\\
		LicenseManager & 296.1 & 0.3 & 187.6 & 108.1\\
		OTP & 170.0 & 0.9 & \ 69.2 & \ 99.9\\
		\bottomrule
	\end{tabular}
     }
    \caption{ESP times in seconds.}
	\label{tab:adss-times}
\end{table}

Our evaluation also includes a purely technical analysis of the performance of the algorithms and techniques used in the \esp on both the reference use cases and artificial applications.

First, we have measured the execution time of the final version of the \esp on the three use cases, with the assets annotated by the experts, as reported in Table~\ref{tab:useCase}.
In all three cases, 17 PIs have been considered using the {\softprot}s listed in Table~\ref{tab:protections}. Opaque predicates, branch functions, and control flow flattening were applied at three configuration levels (low, medium, and high frequencies with corresponding overhead levels). Data obfuscation included three techniques: XOR-masking, residue number encoding, and data-to-procedural conversions~\cite{collberg1997taxonomy}.

Table~\ref{tab:adss-times} shows the \esp computation times. The framing phase is almost instantaneous and driven by the lines of annotated code. Regarding the assessment and the mitigation phases, these measurements do not provide sufficient data for a full assessment of the scalability and complexity of the used algorithms, \changed{as the three applications do not provide enough data points to identify correlations between the computation times and the number of assets/PIs.}

We hence complemented the measurements with a formal evaluation of the algorithms' complexity and a performance measurement on artificial scenarios (in FEDS terminology).
The formal validation investigated the most influential factors for the attack discovery tool and the game-theoretic optimization of the mitigation phase. The complexity of the attack discovery algorithms is exponential in the number of attack steps in the \kb, hence the need to consider pruning strategies.
The complexity of the algorithms in the mitigation phase is linear in the number of assets. It exponentially depends on the number of PIs and the number of attacks discovered in the assessment phase. In this case, having a limited number of PIs and pruning the sequences of {\softprot}s helped with reasonable performance.

To assess scalability, we evaluated the performance of the \esp on three synthetic standalone Linux applications with an increasing number of assets. Table~\ref{tab:expUseCase} summarizes their metrics.
These artificial applications have been randomly generated with a process that selects a call graph (from a set of call graphs extracted from real applications), and then generates randomized function bodies to meet specific code metrics. Then it randomly selects fragments in the generated code as data or code assets.
In this experiment, we used all the previously listed PIs except white-box crypto (which was a proprietary algorithm of one industrial ASPIRE partner). We also added four instances of obfuscation using Tigress, i.e., the ones marked in Table~\ref{tab:protections}.

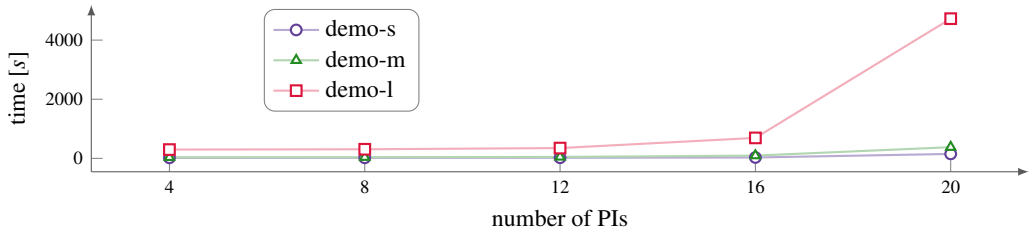
\begin{figure}[t]
	\centering
	\small
	\begin{tikzpicture}
	\begin{axis}
	[legend style={at={(0.35,0.975)}, legend cell align={left}, anchor=north east, draw=gray, rounded corners}, height=12em, width=0.95\linewidth, axis y line*=left, axis x line*=bottom, axis line style={Black!65, -latex}, xlabel={number of PIs}, ylabel={time [$s$]}, xtick={4,8,12,16,20}]
	\addplot[RoyalPurple!35, thick, mark = *, mark options={solid, RoyalPurple, fill=white}, thick] coordinates {(4,17.399) (8,18.129) (12,20.27) (16,29.674) (20,149.757)};
	\addplot[ForestGreen!35, thick, mark = triangle*, mark options={solid, ForestGreen, fill=white}, thick] coordinates {(4,35.489) (8,38.726) (12,48.274) (16,93.432) (20,377.8)};
	\addplot[Crimson!35, thick, mark = square*, mark options={solid, Crimson, fill=white}, thick] coordinates {(4,297.485) (8,307.439) (12,348.038) (16,691.854) (20,4724.334)};
	\legend{demo-s, demo-m, demo-l} 
	\end{axis}
	\end{tikzpicture}
	\caption{\esp execution times on applications reported in Table~\ref{tab:expUseCase}.}
	\label{fig:espTime}
\end{figure}

\changed{On the three artificial applications, we deployed the asset protection optimization approach described in Section~\ref{sec:esp:risk_mitigation} multiple times for different configurations that feature varying numbers of available PIs.}
This deployment was done on an Intel i7-8750H workstation with 32~\si{GB} RAM, using Java 1.8.0\_212 under GNU/Linux Debian 4.18.0. 
Figure~\ref{fig:espTime} depicts the measured total \esp computation time, along with the time needed for the risk assessment, asset protection, and asset hiding phases.
The time needed to complete the workflow increases with the number of PIs considered during the mitigation; such an increase strongly depends on the application code complexity, and in particular on its \sloc and number of assets and functions.

The time needed to analyze the applications' source code and to generate the application meta-model instance was negligible at less than 1\si{s}. The time required to deploy the solution is irrelevant for assessing {the} \esp' computational feasibility, as it only measures the negligible time needed to execute the external {\softprot} tools for the single selected solution.

As expected, the time needed to execute the risk assessment phase does not depend on the number of PIs available to protect the application, as attacks are determined on the vanilla application. Nonetheless, we report that it has limited impact because of the aggressive pruning we have implemented that avoids exponential growth.
The asset protection phase is by far the most computationally intensive, especially when the number of available PIs increases. 
Since the mitigation considers sequences of {\softprot}s, the execution time is exponential as it depends on the combination of PIs. The same holds for the asset hiding phase, although less time is needed to execute the latter compared to the asset protection phase.

These experiments allowed the positive evaluation of \changed{RQ1.d.1, as the computation times were considered acceptable by both the tool developers and the involved experts. They also enable the positive evaluation of RQ1.d.2, as the heuristics implemented in the game-theoretic approach scaled sufficiently well in our experiments to allow producing solutions in useful time, even though its theoretic worst-case behavior might be intractable.} 

\begin{table}[t]
	\centering
	{
	\small
	\begin{tabular}{lrrrrr}
		\toprule
		\multirow{2}{*}{\textsc{Application}} & \multirow{2}{*}{\textsc{\sloc}} & \multirow{2}{*}{\textsc{functions}} & \multicolumn{3}{c}{\textsc{assets}}\\
		\cmidrule(lr){4-6}
		&  &  & \textsc{code} & \textsc{data} & \textsc{total}\\
		\midrule
		demo-s & 443 & 18 & 2 & 2 & 4 \\
		demo-m & 1029 & 47 & 12 & 3 & 15 \\
		demo-l & 3749 & 178 & 26 & 13 & 39 \\
		\bottomrule
	\end{tabular}
	}
	\caption{Statistics of \acrshort{esp} experimental assessment applications.}
	\label{tab:expUseCase}
\end{table}

\subsection{Framing Effort in the ESP}
\label{sec:time}
As mentioned in Section~\ref{sec:external_expert_design}, all framing tasks, including the annotation of the assets in the source code, were performed before the external industrial experts got involved in the assessment of the ESP. This enabled them to focus on the qualitative assessment without wasting their precious time on the more mechanistic framing tasks. This section analyzes the effort that is needed to perform those framing tasks with the ESP.

First, the user needs to annotate the assets in the source code. The ESP supports two options: manually annotating the source code or manually tagging code and data elements in the \esp. Our experiments only used source code annotations, which consist of (mostly single-line) pragmas and attributes~\cite{D5.11,D5.13} that identify the assets and specify security requirements. For users proficient with their syntax, typing out the annotations requires at most tens of seconds per asset.  

More time is required to determine precisely which elements in the source code need to be annotated because they correspond to the application assets. Security architects and software designers describe assets abstractly. When those are known upfront, developers can annotate their code as they write it, thus only requiring the aforementioned tens of seconds per asset. 
For developers that are familiar with the code base but need to annotate the code afterwards, we estimate that locating the assets in the code takes less than a minute for assets with high locality (e.g., single variables or single functions) to potentially tens of minutes for assets that are spread out more throughout the code base (e.g., the invocations of a specific reaction mechanism spread throughout the code for remote attestation).

The annotations were added to the ASPIRE use cases by their original developers after the development was finished. It was the first time they were adding our style of annotations. They hence faced a learning curve. Moreover, while adding the annotations, they had to validate the syntax and expressiveness of the annotation language on the fly. Had they already been proficient with the annotations beforehand and had they just needed to inject them without having to validate their design, we estimate that the time needed to annotate their use cases would have been less than one hour for the use cases with 25 assets, and less than two hours for the one with 43 assets.

In any case, locating the assets in the code base given abstract descriptions is something that any user of any non-trivial \softprot tool needs to do, both to configure the tool to protect the relevant code and to validate that that code has actually been protected by the tool. 
So compared to other \softprot approaches, the mentioned times are not considered overhead required to use the ESP. 
The same holds for selecting the attacks the user wants to mitigate and for determining which of the available {\softprot}s to consider. In the ESP, selecting attacks and {\softprot}s from the ones modelled in the KB happens with a click-of-a-button GUI interface. The time required for clicking is negligible compared to the time for deciding which ones to include or exclude. That decision making needs to happen with any decision support tool, so the ESP is not less efficient in this regard than any other decision making process. 
This discussion completes the answer to RQ1.d.1.

\subsection{Threats to Validity}

We have checked our \changed{evaluation procedure} against a checklist of the possible threats to validity: construct, internal, conclusion, and external validity threats~\cite{wohlin00}, as well as instantiation validity threats~\cite{lukyanenko2014instantiation}.
	
Threats to \emph{construct validity} concern the metrics defined for the experiment.
We have used a set of standard metrics from the ISO. \changed{Nonetheless, the risk remains that the selected metrics are not the best ones for our assessment.} The evaluation scores were positive, negative, and partial, which appeared expressive enough for our purposes. However, we could not objectively assess these criteria' satisfaction as the questionnaires included open answers. 
To limit subjectivity, we have evaluated the answers within the ASPIRE project. Moreover, the ASPIRE project reviewers hired by the European Commission did not contradict our conclusions.
 
Threats to \emph{internal validity} concern the inferences between independent variables and experiment outcomes.
\changed{One possible noise factor that may confound the inference relates to the task comprehension. We assess this factor as negligible in this case.}
The task, resembling their day-to-day job and their typical applications, was described by the experts as clear; the use of the tool was documented; moreover, the experts were assisted in case of doubts (by their colleagues or by us). 
\changed{Another potential noise factor is the experts' commitment to do their tasks diligently before answering the questionnaire.}
We gave the experts the tool and the reports to be assessed offline. 
We hence cannot establish the effort they invested in using and reading them. 
We checked that all relevant artifacts were analysed in their comments (main attack paths and all the combinations of protections); however, we cannot assess their commitment accurately.
\changed{Moreover, another confounding factor is the actual objective evaluation.
The experts were selected from the industrial partners of the project. Even if they were asked to evaluate the artifact objectively, their judgment may have been biased by the will not to hinder the project and the positive evaluation by the European funding agency.}

Threats to \emph{external validity} affect the generalisation of the research results to the real world, \ie experts who want to protect real applications using an automated decision support system.
\changed{In our case, the subjects of the evaluation are real experts that are protecting apps. The evaluation could hence be generalized to experts with a similar background protecting programs analogous to the ones presented in the evaluation and not too dissimilar from the ones they protect during normal job tasks in the same companies.
However, it does not necessarily extend to other experts protecting different applications in other companies.
Significant effort was invested in the use case applications to ensure that they are representative (in terms of size and complexity) of the code bases such experts have to protect in their daily jobs. However, since every expert only evaluated the \esp on one application, we cannot be sure that applications with different structures or from a different domain would yield similar results.}

\changed{Another potential threat to external validity concerns the possibility to generalize the evaluation made on a specific tool, the \esp, to general \softprot tasks, which may affect answers to RQ3.
In this case, we assessed this threat as negligible. Having automated a specific \softprot task, we have proved that automation is feasible, even if the same task could be done in different, possibly better ways. Moreover, the external experts performing the evaluation did not assess the ESP within the ASPIRE scope and requirements, as they did not know about the ASPIRE-defined scope. Instead, they evaluated it vis-à-vis their day-to-day job requirements. 
The limitations identified in Section~\ref{sec:coverage} on the \esp, which does not automate all tasks, do not apply, as RQ3 is related to individual parts of a risk management approach for \softprot.
}

Threats to \emph{conclusion validity} affect the validity of the methods to draw conclusions from the assessment.
In this case, we have asked experts to answer open questions from a standard questionnaire structured according to the main protection workflow. Given the experts' limited availability, a state-of-the-art controlled experiment was impossible. 
The main concern for our assessment is related to the limited number of experts involved, which does not allow us to use standard statistical methods.



\changed{Finally, threats to \emph{instantiation validity} affect the possibility of considering the artifact we have implemented, i.e., the \esp, an instance of the theoretical object we had in mind, i.e., a semi-automated decision support system for \softprot \cite{instantiation-validity}.
The instantiation space is very large, as one can imagine many ways to implement all the tasks the \esp performs. In this space, we opted for a NIST-based four-phases approach. We have not considered alternative approaches, first and foremost because we were convinced upfront that the standard approach that works for other fields is also valid for \softprot. 
Secondly, we have implemented the \esp in the ASPIRE project, where resources for PoC and completion times were constrained. This relates to another threat, the artifact cost, that prevented the implementation of more alternatives.

Two more threats to instantiation validity apply to the \esp: auxiliary features and emergent properties, which stress the complexity of IT tools and oblige considering additional aspects that are not the main focus of the instantiation.
Indeed, the integration of the components and the definition of the workflow has revealed several auxiliary features related to the UI comprehension, the user experience, and the effectiveness of the designed workflow to cope with daily \softprot experts' tasks. 
Furthermore, several emergent properties appeared related to the complexity of the data, their relationships, and the correct data presentation. 
We tried to mitigate these threats by continuously interacting with the internal experts during the design, development, integration, and validation of the \esp and the data artifact it produces. Moreover, every time we received suggestions in the answers to the questionnaires, we have incorporated them before the next evaluation phase. Nonetheless, we cannot exclude that these threats to the instantiation validity may have an impact.}

%% file: conclusion.tex
\section{Conclusion and future work}
\label{sec:conclusion}

\subsection{Conclusions}


We discussed the necessity and potential benefits of a standardized, formalized, and automated approach for risk management in the context of software protection against man-at-the-end attacks. To that end, we discussed just such a risk management approach for software protections, which we based on the NIST SP800-39 standard for risk management for information security. 

To provide an answer to RQ1 on the feasibility of automated decision support tools to improve the useability of \softprot tools, we developed and presented the \esp design and an evaluation of its \poc implementation. \changed{We found that many human expert judgment tasks can already benefit from automated tools and the data they produce, which experts found sufficiently usable, acceptable, and efficient; and of which they assessed the results as sufficiently correct and comprehensible.}

As an answer to RQ2 on how standardized risk management approaches can be adopted for \softprot, we discussed in detail how the different aspects of software protection deployment decision making could and should be mapped onto risk framing, risk assessment, risk mitigation, and risk monitoring phases. \changed{For all phases combined, we identified 50 required constructs, 6 models, and 32 methods that the adopted approach should entail.}

We answered RQ3 on the feasibility of formalizing and automate parts of the adopted approach by providing a mapping of the abstract construct, model, and method artifacts identified for the adopted approach onto the concrete instantiation artifacts that make up the \esp.

With these answers, we have provided convincing evidence that the proposed approach is feasible and can be automated to a large degree, and deserves the launch of a structured community effort that leads to future standardization and automation.

\subsection{Future Work}

It is clear that quite some future work is needed, however, for the standardization itself, as well as for improving, refining, extending, replacing, and complementing the rather embryonic instantiations of the necessary constructs, models and methods currently available in the presented \esp in support of the automation of tasks in the approach. The artifacts in Tables~\ref{tab:constructs},~\ref{tab:models}, and~\ref{tab:methods} which have no counterpart in the \esp yet are clear examples of where more research is needed.

\changed{Section~\ref{sec:requirements} also highlighted} a number of topics for future work in the form of open issues and open research questions, research directions that we consider interesting, and development steps for which a community effort is needed. Tables~\ref{tab:openissues},~\ref{tab:researchdirs}, and~\ref{tab:futurework} provide an overview of that future work.

\begin{table}[t]
\begin{center}

{\footnotesize
\begin{tabular}{r l}
No. & Open Issue / Research Question Subjects\\
\hline
\openissuelab{1}  & \hyperlink{openissue.1}{secondary asset (a.k.a. pivots, hooks, and mileposts) model design}  \\
\openissuelab{2}  & \hyperlink{openissue.2}{selection and models of protection policy requirements}     \\
\openissuelab{3}  & \hyperlink{openissue.3}{level of detail needed in models of attacker capabilities}   \\
\openissuelab{4}  & \hyperlink{openissue.4}{to what extent worst-case assumptions are useful}      \\
\openissuelab{5}  & \hyperlink{openissue.5}{best abstractions to model software features that impact the execution of attack steps} \\
\openissuelab{6}  & \hyperlink{openissue.6}{empirical validation of such models and metrics, in particular for manual human activities}    \\
\openissuelab{7}  & \hyperlink{openissue.7}{identification of viable attack paths on the basis of software analysis results}   \\
\openissuelab{8}  & \hyperlink{openissue.8}{estimation of attack step's required effort and likelihood of success}  \\
\openissuelab{9}  & \hyperlink{openissue.9}{extent to which automated techniques can replace human pen testing} \\
\openissuelab{10}  & \hyperlink{openissue.10}{required granularity of attack steps forming attack paths} \\
\openissuelab{11}  & \hyperlink{openissue.11}{incorporation of informal information obtained from experts (e.g., pen testers) in automated threat analysis}   \\
\openissuelab{12}  & \hyperlink{openissue.12}{incremental attack path enumeration}    \\
\openissuelab{13} &  \hyperlink{openissue.13}{required precision of pre-deployment \softprot impact estimation}   \\
\openissuelab{14}  & \hyperlink{openissue.14}{pre-deployment potency, resilience, and stealth estimation for layered {\softprot}s}  \\
\openissuelab{15}  & \hyperlink{openissue.15}{pre-deployment estimation of \softprot impact on attack success probability }   \\
\openissuelab{16}  & \hyperlink{openissue.16}{validation of deployed \softprot against assumptions made pre-deployment }\\
\end{tabular}
}

\caption{Open issues and topics of open research questions identified in the paper}
\label{tab:openissues}
\end{center}
\end{table}

\begin{table}[t]
\begin{center}

{\footnotesize
\begin{tabular}{r l}
No. & Research Direction\\
\hline
\researchdirlab{1}  & \hyperlink{researchdir.1}{the concept and use of protection policy requirements} \\
\researchdirlab{2}  & \hyperlink{researchdir.2}{machine learning techniques to identify and quantify feasible attack paths} \\
\researchdirlab{3}  & \hyperlink{researchdir.3}{adoption of exploit generation techniques to identify feasible attack paths}  \\
\researchdirlab{4}  & \hyperlink{researchdir.4}{gradual path from a mostly manual process to automated feasible attack path identification} \\
\researchdirlab{5}  & \hyperlink{researchdir.5}{adoption of risk monetisation to evaluate and prioritize actual risks} \\
\researchdirlab{6}  & \hyperlink{researchdir.6}{adoption of the OWASP risk rating methodology to evaluate and prioritize actual risks}  \\
\researchdirlab{7}  & \hyperlink{researchdir.7}{single-pass selection of layered {\softprot}s with accurate assessment of impact on threats and risks} \\
\researchdirlab{8}  & \hyperlink{researchdir.8}{multi-pass selection of layered {\softprot}s with accurate assessment of impact on threats and risks}\\
\researchdirlab{9}  & \hyperlink{researchdir.9}{machine learning techniques to select the most effective layered combinations of {\softprot}s} \\
\end{tabular}
}

\caption{Potentially interesting research directions identified in the paper}
\label{tab:researchdirs}
\end{center}
\end{table}

\begin{table}[t]
\begin{center}

{\footnotesize
\begin{tabular}{r l}
No. & Required community efforts \\
\hline
\futureworklab{1}  & \hyperlink{futurework.1}{provisioning a complete vocabulary and methodology to describe the risk frame} \\
\futureworklab{2}  & \hyperlink{futurework.2}{provisioning a standard taxonomy of assets and their relevant features}   \\
\futureworklab{3}  & \hyperlink{futurework.3}{provisioning a standard taxonomy of \softprot security requirements} \\
\futureworklab{4}  & \hyperlink{futurework.4}{provisioning and maintaining a living catalog of potential attack steps and their relevant features}  \\
\futureworklab{5}  & \hyperlink{futurework.5}{provisioning a standard taxonomy of SPs and their relevant features in support of decision support} \\
\futureworklab{6}  & \hyperlink{futurework.6}{standardizing methodology for defining the actual threat model, attack surface and attack vectors} \\
\end{tabular}
}

\caption{Topics requiring a collaborative development effort by various stakeholders in the \softprot community}
\label{tab:futurework}
\end{center}
\end{table}